\documentclass[reprint,amsmath,amssymb,aps,prl,groupedaddress,nofootinbib,twocolumn,superscriptaddress]{revtex4-2}
\usepackage{graphicx}
\usepackage{amsthm,amssymb,amsmath,braket,mathdots}
\usepackage{bm}\usepackage[pagebackref=false,pdfnewwindow=true]{hyperref} %\hypersetup{draft}
\usepackage{epstopdf,psfrag}
\usepackage{relsize,amsbsy}
\usepackage[export]{adjustbox}

\usepackage{graphicx,xcolor,tikz}

\usepackage{units} %\nicefrac,units
\usepackage{dsfont} %write unity 1

\DeclareMathOperator{\re}{Re}
\DeclareMathOperator{\im}{Im}

\newcommand{\ii}{{\mathrm{i}}} %imaginary unit is straight
\newcommand{\e}{{\mathrm{e}}} %Euler's number straight
\renewcommand{\v}[1]{\bm{#1}} %write vectors fat

\newcommand{\RuCl}{$\alpha\text{\ensuremath{-}}{\text{RuCl}}_{3}$ }
\newcommand{\dd}{\partial}
\renewcommand{\d}{\mathrm{d}}
\newcommand{\1}{\mathds{1}} %identity matrix

\definecolor{bananayellow}{rgb}{1.0, 0.88, 0.21}
\definecolor{straw}{rgb}{0.32, 0.28, 0.1}

\begin{document}
%\title{Anomalous Quantum Oscillations in $\alpha\text{\ensuremath{-}}{\text{RuCl}}_{3}$ / Graphene Heterostructures}
\title{Anomalous Quantum Oscillations in a Heterostructure of\\ Graphene on a Proximate Quantum Spin Liquid}
\author{V. Leeb}
 	\affiliation{Department of Physics TQM, Technische Universit{\"a}t M{\"u}nchen, James-Franck-Stra{\ss}e 1, D-85748 Garching, Germany}
\author{K. Polyudov}
\affiliation{Max-Planck-Institut f{\"u}r Festk{\"o}rperforschung, Heisenbergstrasse 1, D-70569 Stuttgart, Germany}
\author{S. Mashhadi}
\affiliation{Max-Planck-Institut f{\"u}r Festk{\"o}rperforschung, Heisenbergstrasse 1, D-70569 Stuttgart, Germany}
\author{S. Biswas}
\affiliation{Institut f{\"u}r Theoretische Physik, Goethe-Universit{\"a}t Frankfurt, 60438 Frankfurt am Main, Germany}
\author{Roser Valent\'i}
\affiliation{Institut f{\"u}r Theoretische Physik, Goethe-Universit{\"a}t Frankfurt, 60438 Frankfurt am Main, Germany}
\author{M. Burghard}
\affiliation{Max-Planck-Institut f{\"u}r Festk{\"o}rperforschung, Heisenbergstrasse 1, D-70569 Stuttgart, Germany}
\author{J. Knolle}
 	\affiliation{Department of Physics TQM, Technische Universit{\"a}t M{\"u}nchen, James-Franck-Stra{\ss}e 1, D-85748 Garching, Germany}
 	\affiliation{Munich Center for Quantum Science and Technology (MCQST), 80799 Munich, Germany}
 	\affiliation{\small Blackett Laboratory, Imperial College London, London SW7 2AZ, United Kingdom}
\date{\today}

\begin{abstract}
The quasi two-dimensional Mott insulator $\alpha\text{\ensuremath{-}}{\text{RuCl}}_{3}$ is proximate to the sought-after Kitaev quantum spin liquid (QSL). In a layer of $\alpha\text{\ensuremath{-}}{\text{RuCl}}_{3}$ on graphene the dominant Kitaev exchange is further enhanced by strain. Recently, quantum oscillation (QO) measurements of such $\alpha\text{\ensuremath{-}}{\text{RuCl}}_{3}$ / graphene heterostructures showed an anomalous temperature dependence beyond the standard Lifshitz-Kosevich description. Here, we develop a theory of {\it anomalous QO} in an effective Kitaev-Kondo lattice model in which the itinerant electrons of the graphene layer interact with the correlated magnetic layer via spin interactions. At low temperatures a heavy Fermi liquid emerges such that the neutral Majorana fermion excitations of the Kitaev QSL acquire charge by hybridising with the graphene Dirac band. Using ab-initio calculations to determine the parameters of our low energy model we provide a microscopic theory of {\it anomalous QOs} with a non-LK temperature dependence consistent with our measurements. We show how remnants of fractionalized spin excitations can give rise to characteristic signatures in QO experiments.  
\end{abstract}
	
\maketitle

{\bf \textit{Introduction.}}
Quantum oscillation measurements are a standard tool for determining the electronic structure of metallic materials. Famously, Onsager showed how the oscillation frequency of the magnetization or resistivity as a function of inverse magnetic field are directly related to a metal's Fermi surface~\cite{onsager1952interpretation}. The standard QO theory was then completed in 1956 by Lifshitz and Kosevich who derived their well-known LK-formula~\cite{lifshitz1956theory} for the temperature dependence of the amplitude decay which permits an extraction of the effective mass. Over the following decades, one by one all elementary metals followed this canonical description~\cite{shoenberg2009magnetic}. Even strongly correlated systems like heavy fermion metals~\cite{taillefer1987direct} and cuprate high temperature superconductors were no exceptions, i.e. their QO amplitudes showed LK-behaviour and the main sign of correlation effects are effective mass enhancements~\cite{doiron2007quantum}. 

The observation of QOs with a non-LK temperature dependence in the correlated insulator SmB$_6$~\cite{tan2015unconventional,hartstein2018fermi} came totally unexpected, challenging the canonical description of QO. It initiated the search for QO in other correlated insulators, i.e. YbB$_{12}$~\cite{liu2018fermi,xiang2018quantum}, and motivated theoretical works~\cite{Knolle_Cooper_QOwithoutFS,shen2018quantum,erten2016kondo,baskaran2015majorana,chowdhury2018mixed,sodemann2018quantum,zhang2016quantum} which unearthed scenarios beyond the LK paradigm. For example, inverted band insulators could lead to such {\it anomalous QO} with a non-LK temperature evolution~\cite{Knolle_Cooper_QOwithoutFS}, which were then predicted~\cite{knolle2017anomalous} and subsequently observed in quantum well heterostructures~\cite{han2019anomalous,xiao2019anomalous}. Alternatively, it was shown that charge neutral fermions from fractionalization in strongly correlated insulators can potentially give rise to QOs via indirectly coupling to the orbital magnetic field~\cite{chowdhury2018mixed,sodemann2018quantum}. The latter idea was originally introduced for QSLs~\cite{motrunich2006orbital}, which are quantum disordered long-range entangled magnetic phases~\cite{knolle2019field,savary2016quantum}. While the debate about SmB$_6$ is not settled, an exciting development is the advent of heterostructures with two-dimensional magnets~\cite{burch2018magnetism}, which pave the way for studying novel QO phenomena from the interplay of magnetic fluctuations and itinerant charges.

\begin{figure}[t]
		\centering
		\includegraphics[width=1.0\columnwidth]{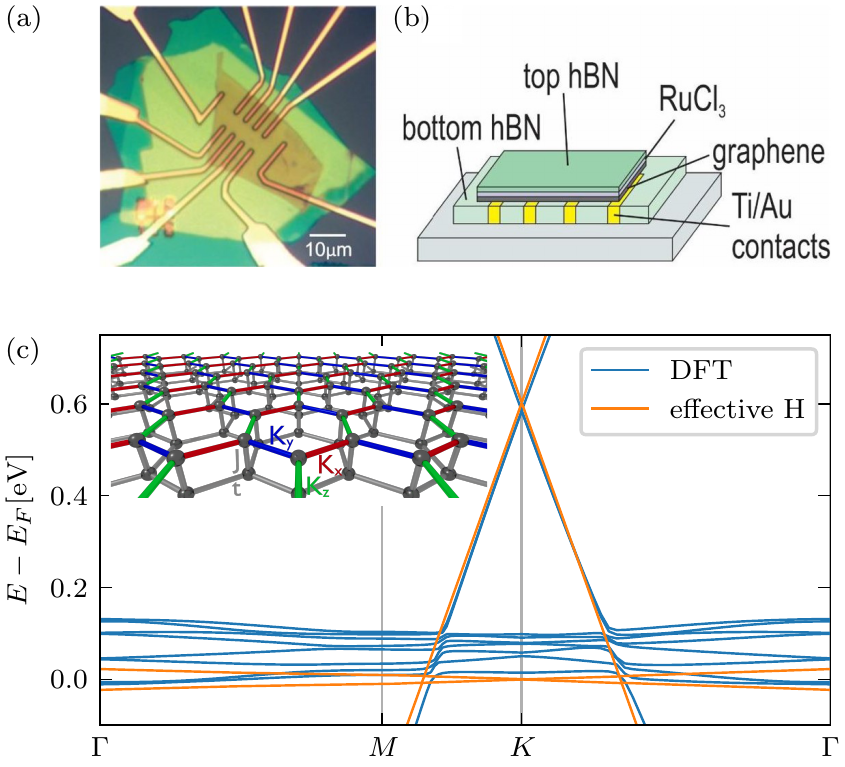}
	\caption{(a) Optical image of a typical $\alpha\text{\ensuremath{-}}{\text{RuCl}}_{3}$ / graphene sample. (b) Schematic illustration of the device, consisting of a flake of hexagonal boron nitride (hBN) as top stacking basis (green), an $\alpha\text{\ensuremath{-}}{\text{RuCl}}_{3}$ nanosheet (purple) and a graphene monolayer (grey). The stack is then transferred onto gold electrodes which are integrated within the bottom hBN flake. (c) Comparison of the electronic structure from ab-initio calculations~\cite{DFT} and the effective low energy model Eq.\eqref{H_lin}. Inset: Schematic picture of the Kitaev-Kondo lattice and its coupling constants, whereby we considered $K^x=K^y=K^z\equiv K$.}
	\label{bandstructure}
\end{figure}

Heterostructures of $\alpha\text{\ensuremath{-}}{\text{RuCl}}_{3}$ on graphene have been recently synthesized~\cite{zhou2019evidence,Mashhadi2019}, see Fig.\ref{bandstructure}(a) and (b), which caused considerable excitement because the insulating $\alpha\text{\ensuremath{-}}{\text{RuCl}}_{3}$ layer is believed to be in proximity to a QSL described by the seminal Kitaev honeycomb model~\cite{Kitaev2006,rau2016spin,hermanns2018physics,Sandilands2015,Banerjee2016,Kasahara2018,winter2017models,takagi2019concept}. The heterostructure undergoes hole (electron) doping of the graphene (Ru) layer~\cite{zhou2019evidence,Mashhadi2019} due to a charge transfer between the correlated insulating layer and itinerant graphene~\cite{DFT,wang2020modulation}. The lattice-mismatch-induced strain is expected to increase the relevance of the Kitaev spin exchange in the magnetic layer~\cite{DFT,gerber2020ab} bringing the system closer to the Kitaev QSL. Intriguingly, resistivity measurements as a function of magnetic field show {\it anomalous QO} with a non-LK temperature dependence. Instead of a monotonic decay as a function of increasing temperature, a maximum appears at a temperature of around 7 K which has been suggested to originate from the magnetic fluctuations of the Kitaev magnet~\cite{Mashhadi2019}.    

Here, we develop a microscopic theory of {\it anomalous QO} in a Kitaev-graphene heterostructure. We provide new QO measurements on graphene in proximity to $\alpha\text{\ensuremath{-}}{\text{RuCl}}_{3}$ and show that our quantitative theory is consistent with their distinct non-LK behaviour. Thus, we establish that the interplay of fractionalized spin excitations and itinerant electrons can lead to {\it anomalous QO} beyond the LK paradigm.  

Our strategy is as follows: We first construct a minimal two-layer model of a Kitaev system coupled to a graphene lattice via spin interactions. Previous works~\cite{Meng2018,choi2018topological} showed within a Majorana mean-field theory (MFT)  that such a Kitaev-Kondo lattice hosts strongly correlated phases like fractionalized Fermi liquids or p-wave superconductivity. In addition, a heavy Fermi-liquid (hFL) phase can be stabilised in which the Majorana fermions of the Kitaev QSL effectively hybridise with the itinerant electrons from the graphene layer. To connect to our measurements, we extract the microscopic parameters of the hFL phase from a tight-binding fit to the ab-initio band structure of Ref.~\cite{DFT}, see Fig.\ref{bandstructure}(c). We then calculate the exact form of the Landau level (LL) structure, which is used to derive an analytic formula for quantum oscillations. 

%%%Alternative Graphic: placing the two subgraphics on top of each other insted of left/right
%\begin{figure}
%	\centering
%	\includegraphics[width=\columnwidth]{Graphics/band_structure_td}
%		\caption{[Version 2 up:down] a) is a schematic picture of the heterostructure where the graphene is shown as a gray lattice, the ruthenium atoms in blue and the chlorine atoms in red. b) is a comparison of the band structure predicted from DFT calculations in \cite{DFT} and the band structure obtained by the linearized Hamiltonian \eqref{H_lin}. The parameters are taken from the fit of the $\unit[20]{nm}$-sample in fig.\ref{fig:R_diffB}.}
%	\label{bandstructure2}
%\end{figure}

{\bf \textit{Effective Low Energy Model.}}
Our starting point is a monolayer Kitaev honeycomb spin model on top of a graphene honeycomb layer (lattice vectors $\v{n_1}$, $\v{n_2}$ and lattice constant $a$). For simplicity we assume that both layers are commensurate but we comment on the effect of lattice mismatch below. The key ingredient is the form of the interaction between the two layers. Because of the Mott insulating nature of $\alpha\text{\ensuremath{-}}{\text{RuCl}}_{3}$ we model it as a spin-only Kondo coupling such that the Hamiltonian $H = H_K+H_t+H_J$ is a sum of three terms:  
\begin{align}
H =& -K \sum_{\langle i j \rangle_\alpha} S^\alpha_i S^\alpha_j -t \sum_{\langle i j \rangle , \sigma}(c^{\dagger}_{i, \sigma} c_{j, \sigma} + h.c.) \nonumber\\
& +  J \sum_{i, \sigma, \sigma', \alpha} c^\dagger_{i,\sigma} \tau^\alpha_{\sigma,\sigma'} c_{i, \sigma'} S^\alpha_i 
\label{H_spinRep}
\end{align}
$S^\alpha$ are the spin-$\frac{1}{2}$ operators (components $\alpha=x,y,z$ also label the inequivalent bond directions $\langle i j \rangle_{\alpha}$ on the honeycomb lattice). The $c^{\dagger}_{i,\sigma}$ create conduction electrons with spin $\sigma$ and $\tau^\alpha$ are the Pauli matrices. 

The exact solution of the Kitaev model proceeds by decomposing the spin operators into real Majorana fermions $\chi_i^\nu$, $S_i^\alpha = \ii \chi_i^0 \chi_i^{\alpha}$~\cite{Kitaev2006}. An important observation is that the ground state flux sector can be described exactly with a Majorana MFT~\cite{Burnell2011,Schaffer2012,knolle2018dynamics}. 
The second term of the Hamiltonian describes the nearest-neighbour hopping with strength $t$ of the electrons. The last part is the Kondo coupling with parameter $J$ between localized spins and itinerant electrons.  

To treat the Kitaev-Kondo lattice we follow previous work~\cite{Meng2018} and rewrite the complex fermions $c_{i, \uparrow}=\frac{1}{\sqrt{2}}(\eta_i^0+i \eta_i^3)$ and $c_{i, \downarrow}=\frac{1}{\sqrt{2}}(i \eta_i^1- \eta_i^2)$  into a sum of Majorana fermions. Then we introduce real mean-fields $U^{\mu \nu}_{i j} = \langle \ii \chi^\mu_i \chi^\nu_j \rangle $ and $W^{\mu \nu}_{i} = \langle \ii \chi^\mu_i \eta^\nu_i \rangle$ such that the MFT Hamiltonian reads (dropping all constants),
\begin{align}
H =& -\frac{K}{4} \sum_{\langle i, j \rangle_\alpha} \ii \v{\chi}_i^T \v{M}^\alpha \v{U}_{i j} \v{M}^\alpha \v{\chi}_j -t \sum_{\langle i, j \rangle} \ii \v{\eta}_i^T \v{\eta}_j +\nonumber\\
&+ \frac{J}{2} \sum_{i, \alpha} \ii \v{\chi}_i^T \v{M}^\alpha \v{W}_{i} \v{M}^\alpha \v{\eta}_i
\label{H_MF}
\end{align}
where the spin matrices $M^\alpha$ are given by $\v{M}^1 = \tau^3 \otimes \ii \tau^2, \v{M}^2 = \ii \tau^2 \otimes \tau^0 \text{ and } \v{M}^3 = \tau^1 \otimes \ii \tau^2$.

In the hFL phase the mean-fields are reduced to $\v{U}=u\1$ and $\v{W}=w\1$~\cite{Meng2018} which allows us to write the Hamiltonian in momentum space in a more familiar form in terms of complex conduction electrons $c_{k, \lambda, \sigma}$ and complex Abrikosov fermions $f_{k, \lambda, \sigma}$ from the Kitaev sector  (with renormalized coupling constants, $K u \rightarrow K$ and $J w \rightarrow J$)
\begin{equation}
H = \sum_{\v{k}, \sigma} 
\begin{pmatrix} c_{\v{k},A,\sigma} \\c_{\v{k},B,\sigma}\\ f_{\v{k},A,\sigma} \\ f_{\v{k},B,\sigma} \end{pmatrix}^\dagger
\begin{pmatrix}
W   & t \theta_{\v{k}} & \frac{J}{2} &    0 \\
t \theta_{\v{k}}^*  & W & 0    & \frac{J}{2} \\
\frac{J}{2} &   0   & 0 & \frac{K}{4} \theta_{\v{k}}  \\
0   & \frac{J}{2}   & \frac{K}{4} \theta_{\v{k}}^* & 0\\
\end{pmatrix}
\begin{pmatrix} c_{\v{k},A,\sigma} \\c_{\v{k},B,\sigma}\\ f_{\v{k},A,\sigma} \\ f_{\v{k},B,\sigma} \end{pmatrix}
\label{eq:H_eff}
\end{equation}
where $\theta_{\v{k}} = 1 + \e^{-\ii \v{k} \v{n_1}} + \e^{-\ii \v{k} \v{n_2}}$ and $\lambda = A, B$ takes account of the two equivalent sublattices. 

The Kondo coupling $J$ now appears as an effective hybridisation between the conduction electrons and the formerly fractionalized fermionic excitations of the Kitaev QSL. Note, our aim is not to solve the MF equations numerically but to connect the hFL band structure to the basic features of the ab-initio electronic structure.

Next, we are interested in the low energy limit and expand $\theta_{K+q}$ linearly around momentum $K$ (the contribution from $K'$ is analogous) to obtain
\begin{align}
H^K = \sum_{\v{q}, \sigma} 
\v{\Phi}_{\v{q},\sigma}^{K \dagger}
\begin{pmatrix}
W \1 + \frac{\sqrt{3}}{2}t a \v{\tau}^*\v{q} & \frac{J}{2} \1 \\
 \frac{J}{2} \1 & \frac{\sqrt{3}}{8}K a \v{\tau}^*\v{q}  \\
\end{pmatrix}
\v{\Phi}_{\v{q},\sigma}^{K}.
\label{H_lin}
\end{align}
The characteristic energy spectrum of $H^K$, see Fig.\ref{bandstructure}(c), consists of a large Dirac cone from the graphene layer which is shifted by $W$ with respect to the smaller and flat Dirac cone of the Kitaev model. We fix the hopping parameter $t=\unit[2.6]{eV}$ by adapting the slope of the graphene Dirac cone to the DFT data. Note, because the Kitaev exchange, $K\approx \unit[17]{meV}$ is much smaller than $t$, the Kitaev subsystem has a strongly reduced bandwidth. The large energy shift $W\approx \unit[0.6]{eV}$ is in accordance with the charge transfer from graphene to \RuCl~\cite{DFT}. Finally, the Fermi energy resides within this correlated layer of the almost flat band, giving rise to the hFL behaviour.  

We note that in \RuCl in proximity to graphene the almost flat Dirac cone originating from the Kitaev QSL is actually gapped due to a small lattice mismatch of the two layers. In principle, this could be simply modelled via a sublattice symmetry breaking term for the Kitaev layer $\sum_{\lambda, \lambda'} f^\dagger_{k,\lambda,\sigma} \tau^z_{\lambda,\lambda'} f_{k,\lambda',\sigma}$, but it does not affect the main results of our work.

%%%LL-Hamiltonian
%\begin{align} 
%&H = \text{const.} \sum_{q_x, \sigma, \zeta = \{K, K'\}}  
%\v{\Psi}_{0,q_x,\sigma}^{\zeta \dagger}
%\begin{pmatrix}
%0& 0 & 0&0\\
%0&W & 0 & \frac{J}{2}\\
%0 &0&0&0 \\
%0&\frac{J}{2} &0& 0\\
%\end{pmatrix}
%\v{\Psi}_{0,q_x,\sigma}^{\zeta} \nonumber\\
%&+ \text{cst.} \sum_{l=1,q_x, \sigma,\zeta}  
%\v{\Psi}_{l,q_x,\sigma}^{\zeta \dagger}
%\begin{pmatrix}
%W & \omega_t \sqrt{l} & \frac{J}{2} &0 \\
%\omega_t \sqrt{l} &W&0&\frac{J}{2} \\
%\frac{J}{2} &0& 0& \omega_K \sqrt{l}\\
%0& \frac{J}{2} & \omega_K \sqrt{l}&0\\
%\end{pmatrix}
%\v{\Psi}_{l,q_x,\sigma}^{\zeta}.
%\label{H_LL}
%\end{align}

{\bf \textit{Landau Levels and Quantum Oscillations.}}
The LLs of the linearized Hamiltonian Eq.\eqref{H_lin} can be found exactly using minimal coupling $\hbar \v{q} \rightarrow \hbar \v{q}- e \v{A}$ with the vector potential $\v{A}$ given in the Landau gauge, see Supplementary Material (SM). For LL index $l>0$ they read
\begin{align}
E_l^{\xi = \pm 1, \zeta = \pm 1} = & \frac{1}{2}\left(\vphantom{\sqrt{\left(W+\xi(\omega_t-\omega_K)\sqrt{l}\right)^2+J^2}}W+\xi(\omega_t+\omega_K) \sqrt{l} \right. \nonumber\\
&\left.+\zeta \sqrt{\left(W+\xi(\omega_t-\omega_K)\sqrt{l}\right)^2+J^2} \right)
\label{LL}
\end{align}
where we defined the cyclotron frequencies $\omega_t = \frac{\sqrt{3}t a}{\sqrt{2} \ell_B}$, $\omega_K = \frac{\sqrt{3}Ka}{4 \sqrt{2} \ell_B}$ and the magnetic length $\ell_B = \sqrt{\frac{\hbar}{e B_z}}$. 
%We have neglected any type of Zeeman-terms.

Each Landau level is $N_\Phi = 4 \frac{B A}{\phi_0}$ fold degenerate, where $A$ is the two-dimensional system size and $\phi_0$ the flux quantum. The factor 4 stems from spin and valley degeneracy at $K$ and $K'$ points. We fix $a=\unit[246]{pm}$ by assuming that the distance between two lattice points is given by the bond-length of graphene \cite{Castro2009}.
Note, using $K\approx \unit[17]{meV}$ from our fit the Kitaev cyclotron frequency $\omega_K$ is with roughly $\unit[50]{\mu eV} \sqrt{B[\mathrm{T}]}$ by far the smallest energy scale.

%\begin{figure}
%	\centering
%	\includegraphics[width=0.8\columnwidth]{Graphics/oscillations_b}
%	\caption{The experimentally determined longitudinal resistance $R_{xx}$ (sample B, blue circles which are connected to guide the eye) and the analytically determined magnetization $M$ of Eq.\eqref{M} are plotted against the cyclotron frequency of graphene $\omega_t \propto \sqrt{B}$. The parameters for $M$ are taken from fitting our low energy model to the ab-initio structure and the amplitude decay in Fig.\ref{fig:R_diffB}.}
%	\label{fig:oscillations}
%\end{figure}
\begin{figure}[t]
			\centering
		\includegraphics[width=1.0\columnwidth]{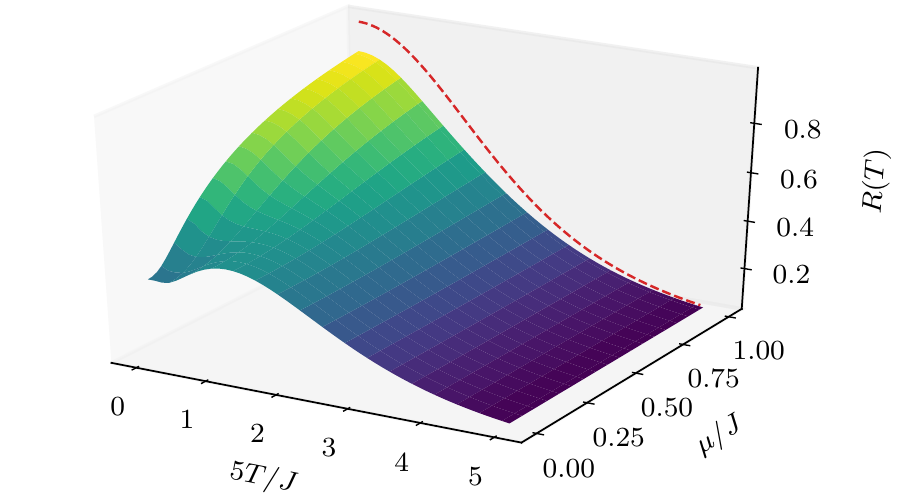}
	\caption{The non-LK behaviour of the QO damping factor $R(T)$ is plotted as a function of temperature and chemical potential deviation $\mu/J$. In the hFL regime for small $\mu/J$ a characteristic maximum appears around $T_{max}\approx J/5$ and at large chemical potential the usual monotonically decreasing LK behaviour (red, dashed line) is recovered.}
	\label{fig:decay}
\end{figure}

Our main objective is to calculate the oscillatory behaviour of observables as a function of magnetic field. As the calculation of transport quantities is cumbersome and requires extra assumptions about scattering channels, we concentrate on thermodynamic quantities directly derived from the grand canonical potential. As the main trend of the frequency and temperature dependence is similar for all observables in metallic systems~\cite{shoenberg2009magnetic}, this allows us to describe the main features of the transport measurements. 

For our analytical calculations we use a connection between the oscillatory part of the grand canonical potential $\Omega_{\text{osc.}}$ and the poles $l^\star_n$ of the finite-temperature Greens function $G^{-1}_{\xi,\zeta}(\ii \omega_n) = \ii \omega_n - (E_l^{\xi,\zeta}-\mu)$ given by 
\begin{equation}
\Omega_{\text{osc.}} = N_\phi T \re \sum_{n=0}^\infty \theta(\re l^\star_n) \sum_{k=1}^\infty \frac{1}{k} \e^{\ii 2 \pi k l^\star_n \mathrm{sgn}\left(\im l^\star_n\right)}.
\label{GLK}
\end{equation} 
This formula was first reported in Ref.~\cite{GLK_Hartnoll_Hofman}, subsequently employed and benchmarked in Ref.~\cite{Knolle_Cooper_QOwithoutFS}, and derived in its general form in our SM. 

Remarkably, in the experimentally relevant limit $\mu,T \ll W$ and $\omega_K\to 0$ the LL structure Eq.\ref{LL} only gives a single pole of the Greens function 
\begin{equation}
l^\star_n = \left(\frac{W}{\omega_t}\right)^2 \left(1-2 \frac{\ii \omega_n}{W} \Gamma \left(\frac{ \mu}{J},\frac{\omega_n}{J}\right)\right),
\label{eq:l_star_simplified}
\end{equation}
 with $\Gamma\left(\frac{\mu}{J},\frac{\omega_n}{J}\right) = 1+ \left(\left(\frac{2 \mu}{J}\right)^2+\left(\frac{2 \omega_n}{J}\right)^2\right)^{-1}$. Note, we have checked that setting $\omega_K= 0$ is consistent with a formal, perturbative expansion in $\nicefrac{\omega_K}{\omega_t}\ll1$. In our approximation we also neglect the term $\frac{\mu}{W} \frac{J^2/4}{ \mu^2 + \omega_n^2}$.

\begin{figure*}[t]
	\centering
	\includegraphics[width=\textwidth]{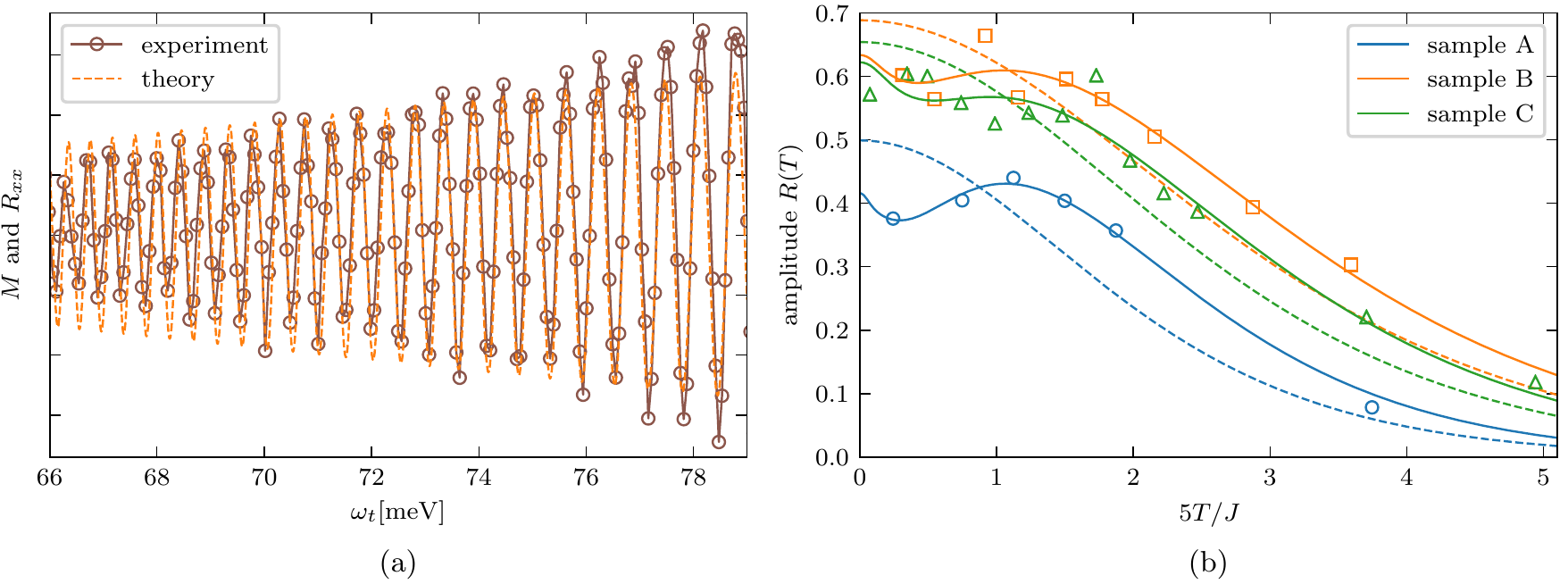}
	%previously \label{fig:oscillations} --> \ref{fig3}~(a)
	%\label{fig:R} --> \ref{fig3}~(b)
	\caption{Panel (a): The experimentally determined longitudinal resistance $R_{xx}$ (sample B at $\unit[3.8]{K}$, circles which are connected to guide the eye) and the analytically determined magnetization $M$ of Eq.\eqref{M} are plotted against the cyclotron frequency of graphene $\omega_t \propto \sqrt{B}$. The parameters for $M$ are taken from the amplitude decay fit in Fig.4 in the SM, with parameters tabulated in Tab.I.
	Panel (b) Fit of the experimental data (open symbols) with the theoretical predicted curves of Eq.(\ref{R}) (solid lines) and the LK damping factor $R_\text{LK} \propto \chi/\sinh \chi$ (dashed lines). Fitting values and quality criteria are given in Tab.I and II in the SM.}
	\label{fig3}
\end{figure*}

As our main result, we then obtain the out-of-plane magnetization from the first harmonic $k=1$ 
\begin{equation}
M = -\frac{\dd \Omega_{\text{osc.}}}{\dd B_z}= - \frac{A W}{\phi_0 \pi} \sin \left( 2 \pi  \left[\frac{W}{\omega_t}\right]^2 \right) R(T).
\label{M}
\end{equation}
Remarkably, for $k=1$ the magnetization is given by a pure sine-oscillation multiplied by a non-LK damping factor
\begin{equation}
R(T) = 2 \chi \sum_{n=0}^\infty \e^{-2 \chi \left(n+\frac{1}{2}\right) \Gamma\left(\frac{ \mu}{J},\frac{\omega_n}{J}\right)} \quad \text{with } \chi = 4 \pi^2 \frac{T W}{\omega_t^2}.
\label{R}
\end{equation}
In contrast to the canonical LK behavior, the damping factor features, for fillings close to the hybridisation region $\mu \sim J$, a maximum at $T \approx \nicefrac{J}{5}$, see Fig.\ref{fig:decay}. 
%Note, higher harmonics can easily be obtained but are in practice usually suppressed by disorder. However, higher harmonics for $M$ lead to broader maxima and sharper minima, which is consistent with experimental data at higher magnetic fields (not shown).

%%wt of 4nm = 96.5 meV
%\begin{figure}[h]
%	\centering
%	\includegraphics[width=0.8\columnwidth]{Graphics/dampingFactor2}
%		\caption{(Color online) Fit of the experimental data (dots) with the theoretical predicted curves (solid lines).}
%	\label{fig:R}
%\end{figure}

% \begin{widetext}
% \begin{eqnarray}
% \label{StructureFactor}
%\mathcal{S}^{\alpha\alpha}_{\mathbf{q}=0}(\omega)  & = &   \sum_n \delta \left[\omega-\Delta-\epsilon^u_n \right]\!\! |X_{n,0}|^2 n_F\left(-\epsilon^u_n/T \right) + \sum_n \delta \left[\omega-\Delta+\epsilon^u_n  \right] |Y_{n,0}|^2 n_F\left(\epsilon^u_n/T \right).
% \end{eqnarray}
%\end{widetext}
{\bf \textit{Comparison to Experiment.} }
We have collected QO data from 3 samples. 
All consist of $\alpha\text{\ensuremath{-}}{\text{RuCl}}_{3}$ flakes of different thickness on a graphene monolayer, see Fig.\ref{bandstructure}(b). For sample A from Ref.~\cite{Mashhadi2019} the thickness is 20 nm, while for the new samples B and C it is 20 nm and 4 nm.
Note, as graphene is only expected to directly interact with the first $\alpha\text{\ensuremath{-}}{\text{RuCl}}_{3}$ layers the thickness variation is not expected to strongly influence the charge transport through the proximitized graphene. The temperature-dependent magnetotransport measurements of the samples were performed in a similar manner as previously described in Ref.~\cite{Mashhadi2019}. In this case we have measured $R_{xx}$ for $\unit[\pm 4]{T}$ to $\unit[\pm 12]{T}$ and various $T$ from $\unit[1]{K}$ up to $\unit[20]{K}$. 

In Fig.\ref{fig3}~(a) we show a typical behaviour of the QO at a fixed temperature for sample B (A and C are very similar). We also show our analytically calculated magnetization which shows the same oscillatory behaviour as the experimentally measured longitudinal resistance $R_{xx}$. We stress that the frequency is in accordance with the charge transfer from ab-initio calculations which was fitted as $W=\unit[603]{meV}$ for our effective model in Fig.\ref{bandstructure}(c). 
%The cyclotron frequency of graphene is $\omega_t = \frac{\sqrt{3}t a}{\sqrt{2} \ell_B}$  with $t=\unit[2.6]{eV}$, such that the slope of the graphene Dirac cone fits nicely to the DFT data. 
To make the comparison of the data to our model quantitative we analysed for each sample the Fourier spectrum as a function of $B^{-1}$ , see Fig.6 in the SM. Each has a sharp maximum in the frequency spectrum which allowed us to extract $W$ independently from Eq.\ref{M}.

Finally, we show the decay of the QO amplitude for increasing temperature in Fig.\ref{fig3}~(b). All three samples have a clear non-LK behaviour with samples A, B displaying a clear maximum around $T_{max}\approx \unit[7]{K}$. We fitted the damping factor Eq.\eqref{R} to the experimental data by varying the parameters $\mu$ and $J$ (for sample B, C at a fixed $B$-field and for sample A from the Fourier transformation of a window around $\unit[10]{T}$). We find $|\mu|<\unit[0.5]{meV}$ and the key parameter is the Kondo exchange $J\approx \unit[2]{meV}$. Details are given in the SM where we also confirm that  we can robustly fit multiple cuts along different $B$-fields without changing parameters $\mu$ and $J$, see Fig.4. We note that, in general, the total amplitude of the anomalous QOs is suppressed exponentially in $\frac{J W}{\omega_t^2}$. However, for our microscopic parameters, in particular our estimates of $J$, this suppression is negligible and the size of anomalous QO is comparable to the standard ones outside of the hybridization region, see SM.

{\bf \textit{Discussion.}}
Our microscopic theory of {\it anomalous QO} is based on the hFL electronic structure of the $\alpha\text{\ensuremath{-}}{\text{RuCl}}_{3}$ / graphene heterostructures and the parameters of the effective model which are determined by ab-initio calculations~\cite{DFT}. The most striking feature of the experimental data is the maximum of the amplitude at a non-zero temperature $T_{max}$ which is reproduced in our theory. 

A stringent test of our scenario would be a controlled change of the interlayer coupling $J$, for example via pressure or intercalation, which should lead to a characteristic shift of the maximum in temperature. Alternatively, because the low energy Kitaev and Kondo scales determine the almost flat band dispersion of the hFL they should be accessible in tunneling measurements. 

Previously, the maximum of the amplitude at $\approx\unit[7]{K}$ has been tentatively interpreted as a signature of the transition to long-range magnetic order in $\alpha\text{\ensuremath{-}}{\text{RuCl}}_{3}$~\cite{Mashhadi2019} because it coincides with its bulk transition temperature $T_N$~\cite{banerjee2018excitations}. However, we argue that this is unlikely because the diverging magnetic fluctuations upon approaching $T_N$ (from higher or lower temperatures) should lead to an {\it increase} in the itinerant electrons scattering rate which then should {\it decrease} the QO amplitude at odds with a maximum. In addition, no direct sign of a magnetic transition has been observed in the heterostructure so far.

{\bf \textit{Summary and Outlook.}}
We have developed a theory of {\it anomalous QO} in $\alpha\text{\ensuremath{-}}{\text{RuCl}}_{3}$ / graphene heterostructures by constructing an effective low energy model which allowed us to derive a new non-LK temperature dependence consistent with our experimental data.

The observation that the frequency and temperature dependence of  {\it anomalous QO} in the Kitaev-Kondo lattice model is consistent with our Shubnikov-de Haas measurements leads to an intriguing interpretation in terms of fractionalized spin excitations within the $\alpha\text{\ensuremath{-}}{\text{RuCl}}_{3}$ layer -- the formerly neutral Majorana fermion excitations of the Kitaev QSL acquire charge via the Kondo coupling to the graphene layer and the ensuing hFL band structure gives rise to the {\it anomalous QO}. Such a scenario directly motivates a search for unconventional superconductivity at lower temperatures which is predicted to occupy a large part of the phase diagram of the  Kitaev-Kondo lattice model~\cite{Meng2018,choi2018topological}. In addition, dynamic fluctuations and collective modes  of these heterostructures, e.g. the recently observed plasmons~\cite{rizzo2020graphene}, are expected to inherit signatures of the proximate QSL.

Our theory of {\it anomalous QO} will be applicable to other systems which can be described as a Dirac semi-metal in contact with a strongly correlated insulator giving rise to an effective heavy band structure with an almost flat band. The formula for the non-LK temperature dependence can then be turned into a versatile tool to extract the low energy scales of the correlated layer. We expect that other magnetic heterostructures, for example with TaS$_2$ films~\cite{law20171t} or the Kagome magnet Nb$_3$X$_8$ (X $=$ Cl, Br)~\cite{pasco2019tunable}, are potential candidates for {\it anomalous QO}. 

On the theory side many questions remain to be explored, for example, how do the fractionalized excitations acquire charge beyond the basic hybridisation picture; or what are alternative microscopic scenarios for QO with charge neutral excitations? Overall, our work paves the way for novel approaches beyond the venerable Lifshitz-Kosevich theory and should also serve as a guide for numerical studies of QO in strongly correlated materials.

\begin{acknowledgments}
\textit{Acknowledgments:} 
JK thanks K. Burch, N.R. Cooper and E. Henriksen for discussions and related collaborations. S.B. and R.V. thank the Deutsche Forschungsgemeinschaft (DFG) for funding through TRR 288-422213477 (A05). 
\end{acknowledgments}

%\bibliography{bib}

\begin{thebibliography}{47}%
\makeatletter
\providecommand \@ifxundefined [1]{%
 \@ifx{#1\undefined}
}%
\providecommand \@ifnum [1]{%
 \ifnum #1\expandafter \@firstoftwo
 \else \expandafter \@secondoftwo
 \fi
}%
\providecommand \@ifx [1]{%
 \ifx #1\expandafter \@firstoftwo
 \else \expandafter \@secondoftwo
 \fi
}%
\providecommand \natexlab [1]{#1}%
\providecommand \enquote  [1]{``#1''}%
\providecommand \bibnamefont  [1]{#1}%
\providecommand \bibfnamefont [1]{#1}%
\providecommand \citenamefont [1]{#1}%
\providecommand \href@noop [0]{\@secondoftwo}%
\providecommand \href [0]{\begingroup \@sanitize@url \@href}%
\providecommand \@href[1]{\@@startlink{#1}\@@href}%
\providecommand \@@href[1]{\endgroup#1\@@endlink}%
\providecommand \@sanitize@url [0]{\catcode `\\12\catcode `\$12\catcode
  `\&12\catcode `\#12\catcode `\^12\catcode `\_12\catcode `\%12\relax}%
\providecommand \@@startlink[1]{}%
\providecommand \@@endlink[0]{}%
\providecommand \url  [0]{\begingroup\@sanitize@url \@url }%
\providecommand \@url [1]{\endgroup\@href {#1}{\urlprefix }}%
\providecommand \urlprefix  [0]{URL }%
\providecommand \Eprint [0]{\href }%
\providecommand \doibase [0]{http://dx.doi.org/}%
\providecommand \selectlanguage [0]{\@gobble}%
\providecommand \bibinfo  [0]{\@secondoftwo}%
\providecommand \bibfield  [0]{\@secondoftwo}%
\providecommand \translation [1]{[#1]}%
\providecommand \BibitemOpen [0]{}%
\providecommand \bibitemStop [0]{}%
\providecommand \bibitemNoStop [0]{.\EOS\space}%
\providecommand \EOS [0]{\spacefactor3000\relax}%
\providecommand \BibitemShut  [1]{\csname bibitem#1\endcsname}%
\let\auto@bib@innerbib\@empty
%</preamble>
\bibitem [{\citenamefont {Onsager}(1952)}]{onsager1952interpretation}%
  \BibitemOpen
  \bibfield  {author} {\bibinfo {author} {\bibfnamefont {L.}~\bibnamefont
  {Onsager}},\ }\href@noop {} {\bibfield  {journal} {\bibinfo  {journal} {The
  London, Edinburgh, and Dublin Philosophical Magazine and Journal of Science}\
  }\textbf {\bibinfo {volume} {43}},\ \bibinfo {pages} {1006} (\bibinfo {year}
  {1952})}\BibitemShut {NoStop}%
\bibitem [{\citenamefont {Lifshitz}\ and\ \citenamefont
  {Kosevich}(1956)}]{lifshitz1956theory}%
  \BibitemOpen
  \bibfield  {author} {\bibinfo {author} {\bibfnamefont {I.}~\bibnamefont
  {Lifshitz}}\ and\ \bibinfo {author} {\bibfnamefont {A.}~\bibnamefont
  {Kosevich}},\ }\href@noop {} {\bibfield  {journal} {\bibinfo  {journal} {Sov.
  Phys. JETP}\ }\textbf {\bibinfo {volume} {2}},\ \bibinfo {pages} {636}
  (\bibinfo {year} {1956})}\BibitemShut {NoStop}%
\bibitem [{\citenamefont {Shoenberg}(2009)}]{shoenberg2009magnetic}%
  \BibitemOpen
  \bibfield  {author} {\bibinfo {author} {\bibfnamefont {D.}~\bibnamefont
  {Shoenberg}},\ }\href@noop {} {\emph {\bibinfo {title} {Magnetic oscillations
  in metals}}}\ (\bibinfo  {publisher} {Cambridge university press},\ \bibinfo
  {year} {2009})\BibitemShut {NoStop}%
\bibitem [{\citenamefont {Taillefer}\ \emph {et~al.}(1987)\citenamefont
  {Taillefer}, \citenamefont {Newbury}, \citenamefont {Lonzarich},
  \citenamefont {Fisk},\ and\ \citenamefont {Smith}}]{taillefer1987direct}%
  \BibitemOpen
  \bibfield  {author} {\bibinfo {author} {\bibfnamefont {L.}~\bibnamefont
  {Taillefer}}, \bibinfo {author} {\bibfnamefont {R.}~\bibnamefont {Newbury}},
  \bibinfo {author} {\bibfnamefont {G.}~\bibnamefont {Lonzarich}}, \bibinfo
  {author} {\bibfnamefont {Z.}~\bibnamefont {Fisk}}, \ and\ \bibinfo {author}
  {\bibfnamefont {J.}~\bibnamefont {Smith}},\ }in\ \href@noop {} {\emph
  {\bibinfo {booktitle} {Anomalous Rare Earths and Actinides}}}\ (\bibinfo
  {publisher} {Elsevier},\ \bibinfo {year} {1987})\ pp.\ \bibinfo {pages}
  {372--376}\BibitemShut {NoStop}%
\bibitem [{\citenamefont {Doiron-Leyraud}\ \emph {et~al.}(2007)\citenamefont
  {Doiron-Leyraud}, \citenamefont {Proust}, \citenamefont {LeBoeuf},
  \citenamefont {Levallois}, \citenamefont {Bonnemaison}, \citenamefont
  {Liang}, \citenamefont {Bonn}, \citenamefont {Hardy},\ and\ \citenamefont
  {Taillefer}}]{doiron2007quantum}%
  \BibitemOpen
  \bibfield  {author} {\bibinfo {author} {\bibfnamefont {N.}~\bibnamefont
  {Doiron-Leyraud}}, \bibinfo {author} {\bibfnamefont {C.}~\bibnamefont
  {Proust}}, \bibinfo {author} {\bibfnamefont {D.}~\bibnamefont {LeBoeuf}},
  \bibinfo {author} {\bibfnamefont {J.}~\bibnamefont {Levallois}}, \bibinfo
  {author} {\bibfnamefont {J.-B.}\ \bibnamefont {Bonnemaison}}, \bibinfo
  {author} {\bibfnamefont {R.}~\bibnamefont {Liang}}, \bibinfo {author}
  {\bibfnamefont {D.}~\bibnamefont {Bonn}}, \bibinfo {author} {\bibfnamefont
  {W.}~\bibnamefont {Hardy}}, \ and\ \bibinfo {author} {\bibfnamefont
  {L.}~\bibnamefont {Taillefer}},\ }\href@noop {} {\bibfield  {journal}
  {\bibinfo  {journal} {Nature}\ }\textbf {\bibinfo {volume} {447}},\ \bibinfo
  {pages} {565} (\bibinfo {year} {2007})}\BibitemShut {NoStop}%
\bibitem [{\citenamefont {Tan}\ \emph {et~al.}(2015)\citenamefont {Tan},
  \citenamefont {Hsu}, \citenamefont {Zeng}, \citenamefont {Hatnean},
  \citenamefont {Harrison}, \citenamefont {Zhu}, \citenamefont {Hartstein},
  \citenamefont {Kiourlappou}, \citenamefont {Srivastava}, \citenamefont
  {Johannes} \emph {et~al.}}]{tan2015unconventional}%
  \BibitemOpen
  \bibfield  {author} {\bibinfo {author} {\bibfnamefont {B.}~\bibnamefont
  {Tan}}, \bibinfo {author} {\bibfnamefont {Y.-T.}\ \bibnamefont {Hsu}},
  \bibinfo {author} {\bibfnamefont {B.}~\bibnamefont {Zeng}}, \bibinfo {author}
  {\bibfnamefont {M.~C.}\ \bibnamefont {Hatnean}}, \bibinfo {author}
  {\bibfnamefont {N.}~\bibnamefont {Harrison}}, \bibinfo {author}
  {\bibfnamefont {Z.}~\bibnamefont {Zhu}}, \bibinfo {author} {\bibfnamefont
  {M.}~\bibnamefont {Hartstein}}, \bibinfo {author} {\bibfnamefont
  {M.}~\bibnamefont {Kiourlappou}}, \bibinfo {author} {\bibfnamefont
  {A.}~\bibnamefont {Srivastava}}, \bibinfo {author} {\bibfnamefont
  {M.}~\bibnamefont {Johannes}},  \emph {et~al.},\ }\href@noop {} {\bibfield
  {journal} {\bibinfo  {journal} {Science}\ }\textbf {\bibinfo {volume}
  {349}},\ \bibinfo {pages} {287} (\bibinfo {year} {2015})}\BibitemShut
  {NoStop}%
\bibitem [{\citenamefont {Hartstein}\ \emph {et~al.}(2018)\citenamefont
  {Hartstein}, \citenamefont {Toews}, \citenamefont {Hsu}, \citenamefont
  {Zeng}, \citenamefont {Chen}, \citenamefont {Hatnean}, \citenamefont {Zhang},
  \citenamefont {Nakamura}, \citenamefont {Padgett}, \citenamefont
  {Rodway-Gant} \emph {et~al.}}]{hartstein2018fermi}%
  \BibitemOpen
  \bibfield  {author} {\bibinfo {author} {\bibfnamefont {M.}~\bibnamefont
  {Hartstein}}, \bibinfo {author} {\bibfnamefont {W.}~\bibnamefont {Toews}},
  \bibinfo {author} {\bibfnamefont {Y.-T.}\ \bibnamefont {Hsu}}, \bibinfo
  {author} {\bibfnamefont {B.}~\bibnamefont {Zeng}}, \bibinfo {author}
  {\bibfnamefont {X.}~\bibnamefont {Chen}}, \bibinfo {author} {\bibfnamefont
  {M.~C.}\ \bibnamefont {Hatnean}}, \bibinfo {author} {\bibfnamefont
  {Q.}~\bibnamefont {Zhang}}, \bibinfo {author} {\bibfnamefont
  {S.}~\bibnamefont {Nakamura}}, \bibinfo {author} {\bibfnamefont
  {A.}~\bibnamefont {Padgett}}, \bibinfo {author} {\bibfnamefont
  {G.}~\bibnamefont {Rodway-Gant}},  \emph {et~al.},\ }\href@noop {} {\bibfield
   {journal} {\bibinfo  {journal} {Nature Physics}\ }\textbf {\bibinfo {volume}
  {14}},\ \bibinfo {pages} {166} (\bibinfo {year} {2018})}\BibitemShut
  {NoStop}%
\bibitem [{\citenamefont {Liu}\ \emph {et~al.}(2018)\citenamefont {Liu},
  \citenamefont {Hartstein}, \citenamefont {Wallace}, \citenamefont {Davies},
  \citenamefont {Hatnean}, \citenamefont {Johannes}, \citenamefont
  {Shitsevalova}, \citenamefont {Balakrishnan},\ and\ \citenamefont
  {Sebastian}}]{liu2018fermi}%
  \BibitemOpen
  \bibfield  {author} {\bibinfo {author} {\bibfnamefont {H.}~\bibnamefont
  {Liu}}, \bibinfo {author} {\bibfnamefont {M.}~\bibnamefont {Hartstein}},
  \bibinfo {author} {\bibfnamefont {G.~J.}\ \bibnamefont {Wallace}}, \bibinfo
  {author} {\bibfnamefont {A.~J.}\ \bibnamefont {Davies}}, \bibinfo {author}
  {\bibfnamefont {M.~C.}\ \bibnamefont {Hatnean}}, \bibinfo {author}
  {\bibfnamefont {M.~D.}\ \bibnamefont {Johannes}}, \bibinfo {author}
  {\bibfnamefont {N.}~\bibnamefont {Shitsevalova}}, \bibinfo {author}
  {\bibfnamefont {G.}~\bibnamefont {Balakrishnan}}, \ and\ \bibinfo {author}
  {\bibfnamefont {S.~E.}\ \bibnamefont {Sebastian}},\ }\href@noop {} {\bibfield
   {journal} {\bibinfo  {journal} {Journal of Physics: Condensed Matter}\
  }\textbf {\bibinfo {volume} {30}},\ \bibinfo {pages} {16LT01} (\bibinfo
  {year} {2018})}\BibitemShut {NoStop}%
\bibitem [{\citenamefont {Xiang}\ \emph {et~al.}(2018)\citenamefont {Xiang},
  \citenamefont {Kasahara}, \citenamefont {Asaba}, \citenamefont {Lawson},
  \citenamefont {Tinsman}, \citenamefont {Chen}, \citenamefont {Sugimoto},
  \citenamefont {Kawaguchi}, \citenamefont {Sato}, \citenamefont {Li} \emph
  {et~al.}}]{xiang2018quantum}%
  \BibitemOpen
  \bibfield  {author} {\bibinfo {author} {\bibfnamefont {Z.}~\bibnamefont
  {Xiang}}, \bibinfo {author} {\bibfnamefont {Y.}~\bibnamefont {Kasahara}},
  \bibinfo {author} {\bibfnamefont {T.}~\bibnamefont {Asaba}}, \bibinfo
  {author} {\bibfnamefont {B.}~\bibnamefont {Lawson}}, \bibinfo {author}
  {\bibfnamefont {C.}~\bibnamefont {Tinsman}}, \bibinfo {author} {\bibfnamefont
  {L.}~\bibnamefont {Chen}}, \bibinfo {author} {\bibfnamefont {K.}~\bibnamefont
  {Sugimoto}}, \bibinfo {author} {\bibfnamefont {S.}~\bibnamefont {Kawaguchi}},
  \bibinfo {author} {\bibfnamefont {Y.}~\bibnamefont {Sato}}, \bibinfo {author}
  {\bibfnamefont {G.}~\bibnamefont {Li}},  \emph {et~al.},\ }\href@noop {}
  {\bibfield  {journal} {\bibinfo  {journal} {Science}\ }\textbf {\bibinfo
  {volume} {362}},\ \bibinfo {pages} {65} (\bibinfo {year} {2018})}\BibitemShut
  {NoStop}%
\bibitem [{\citenamefont {Knolle}\ and\ \citenamefont
  {Cooper}(2015)}]{Knolle_Cooper_QOwithoutFS}%
  \BibitemOpen
  \bibfield  {author} {\bibinfo {author} {\bibfnamefont {J.}~\bibnamefont
  {Knolle}}\ and\ \bibinfo {author} {\bibfnamefont {N.~R.}\ \bibnamefont
  {Cooper}},\ }\href {\doibase 10.1103/PhysRevLett.115.146401} {\bibfield
  {journal} {\bibinfo  {journal} {Phys. Rev. Lett.}\ }\textbf {\bibinfo
  {volume} {115}},\ \bibinfo {pages} {146401} (\bibinfo {year}
  {2015})}\BibitemShut {NoStop}%
\bibitem [{\citenamefont {Shen}\ and\ \citenamefont
  {Fu}(2018)}]{shen2018quantum}%
  \BibitemOpen
  \bibfield  {author} {\bibinfo {author} {\bibfnamefont {H.}~\bibnamefont
  {Shen}}\ and\ \bibinfo {author} {\bibfnamefont {L.}~\bibnamefont {Fu}},\
  }\href@noop {} {\bibfield  {journal} {\bibinfo  {journal} {Physical review
  letters}\ }\textbf {\bibinfo {volume} {121}},\ \bibinfo {pages} {026403}
  (\bibinfo {year} {2018})}\BibitemShut {NoStop}%
\bibitem [{\citenamefont {Erten}\ \emph {et~al.}(2016)\citenamefont {Erten},
  \citenamefont {Ghaemi},\ and\ \citenamefont {Coleman}}]{erten2016kondo}%
  \BibitemOpen
  \bibfield  {author} {\bibinfo {author} {\bibfnamefont {O.}~\bibnamefont
  {Erten}}, \bibinfo {author} {\bibfnamefont {P.}~\bibnamefont {Ghaemi}}, \
  and\ \bibinfo {author} {\bibfnamefont {P.}~\bibnamefont {Coleman}},\
  }\href@noop {} {\bibfield  {journal} {\bibinfo  {journal} {Physical review
  letters}\ }\textbf {\bibinfo {volume} {116}},\ \bibinfo {pages} {046403}
  (\bibinfo {year} {2016})}\BibitemShut {NoStop}%
\bibitem [{\citenamefont {Baskaran}(2015)}]{baskaran2015majorana}%
  \BibitemOpen
  \bibfield  {author} {\bibinfo {author} {\bibfnamefont {G.}~\bibnamefont
  {Baskaran}},\ }\href@noop {} {\bibfield  {journal} {\bibinfo  {journal}
  {arXiv preprint arXiv:1507.03477}\ } (\bibinfo {year} {2015})}\BibitemShut
  {NoStop}%
\bibitem [{\citenamefont {Chowdhury}\ \emph {et~al.}(2018)\citenamefont
  {Chowdhury}, \citenamefont {Sodemann},\ and\ \citenamefont
  {Senthil}}]{chowdhury2018mixed}%
  \BibitemOpen
  \bibfield  {author} {\bibinfo {author} {\bibfnamefont {D.}~\bibnamefont
  {Chowdhury}}, \bibinfo {author} {\bibfnamefont {I.}~\bibnamefont {Sodemann}},
  \ and\ \bibinfo {author} {\bibfnamefont {T.}~\bibnamefont {Senthil}},\
  }\href@noop {} {\bibfield  {journal} {\bibinfo  {journal} {Nature
  communications}\ }\textbf {\bibinfo {volume} {9}},\ \bibinfo {pages} {1}
  (\bibinfo {year} {2018})}\BibitemShut {NoStop}%
\bibitem [{\citenamefont {Sodemann}\ \emph {et~al.}(2018)\citenamefont
  {Sodemann}, \citenamefont {Chowdhury},\ and\ \citenamefont
  {Senthil}}]{sodemann2018quantum}%
  \BibitemOpen
  \bibfield  {author} {\bibinfo {author} {\bibfnamefont {I.}~\bibnamefont
  {Sodemann}}, \bibinfo {author} {\bibfnamefont {D.}~\bibnamefont {Chowdhury}},
  \ and\ \bibinfo {author} {\bibfnamefont {T.}~\bibnamefont {Senthil}},\
  }\href@noop {} {\bibfield  {journal} {\bibinfo  {journal} {Physical Review
  B}\ }\textbf {\bibinfo {volume} {97}},\ \bibinfo {pages} {045152} (\bibinfo
  {year} {2018})}\BibitemShut {NoStop}%
\bibitem [{\citenamefont {Zhang}\ \emph {et~al.}(2016)\citenamefont {Zhang},
  \citenamefont {Song},\ and\ \citenamefont {Wang}}]{zhang2016quantum}%
  \BibitemOpen
  \bibfield  {author} {\bibinfo {author} {\bibfnamefont {L.}~\bibnamefont
  {Zhang}}, \bibinfo {author} {\bibfnamefont {X.-Y.}\ \bibnamefont {Song}}, \
  and\ \bibinfo {author} {\bibfnamefont {F.}~\bibnamefont {Wang}},\ }\href@noop
  {} {\bibfield  {journal} {\bibinfo  {journal} {Physical review letters}\
  }\textbf {\bibinfo {volume} {116}},\ \bibinfo {pages} {046404} (\bibinfo
  {year} {2016})}\BibitemShut {NoStop}%
\bibitem [{\citenamefont {Knolle}\ and\ \citenamefont
  {Cooper}(2017)}]{knolle2017anomalous}%
  \BibitemOpen
  \bibfield  {author} {\bibinfo {author} {\bibfnamefont {J.}~\bibnamefont
  {Knolle}}\ and\ \bibinfo {author} {\bibfnamefont {N.~R.}\ \bibnamefont
  {Cooper}},\ }\href@noop {} {\bibfield  {journal} {\bibinfo  {journal}
  {Physical Review Letters}\ }\textbf {\bibinfo {volume} {118}},\ \bibinfo
  {pages} {176801} (\bibinfo {year} {2017})}\BibitemShut {NoStop}%
\bibitem [{\citenamefont {Han}\ \emph {et~al.}(2019)\citenamefont {Han},
  \citenamefont {Li}, \citenamefont {Zhang}, \citenamefont {Sullivan},\ and\
  \citenamefont {Du}}]{han2019anomalous}%
  \BibitemOpen
  \bibfield  {author} {\bibinfo {author} {\bibfnamefont {Z.}~\bibnamefont
  {Han}}, \bibinfo {author} {\bibfnamefont {T.}~\bibnamefont {Li}}, \bibinfo
  {author} {\bibfnamefont {L.}~\bibnamefont {Zhang}}, \bibinfo {author}
  {\bibfnamefont {G.}~\bibnamefont {Sullivan}}, \ and\ \bibinfo {author}
  {\bibfnamefont {R.-R.}\ \bibnamefont {Du}},\ }\href@noop {} {\bibfield
  {journal} {\bibinfo  {journal} {Physical review letters}\ }\textbf {\bibinfo
  {volume} {123}},\ \bibinfo {pages} {126803} (\bibinfo {year}
  {2019})}\BibitemShut {NoStop}%
\bibitem [{\citenamefont {Xiao}\ \emph {et~al.}(2019)\citenamefont {Xiao},
  \citenamefont {Liu}, \citenamefont {Samarth},\ and\ \citenamefont
  {Hu}}]{xiao2019anomalous}%
  \BibitemOpen
  \bibfield  {author} {\bibinfo {author} {\bibfnamefont {D.}~\bibnamefont
  {Xiao}}, \bibinfo {author} {\bibfnamefont {C.-X.}\ \bibnamefont {Liu}},
  \bibinfo {author} {\bibfnamefont {N.}~\bibnamefont {Samarth}}, \ and\
  \bibinfo {author} {\bibfnamefont {L.-H.}\ \bibnamefont {Hu}},\ }\href@noop {}
  {\bibfield  {journal} {\bibinfo  {journal} {Physical review letters}\
  }\textbf {\bibinfo {volume} {122}},\ \bibinfo {pages} {186802} (\bibinfo
  {year} {2019})}\BibitemShut {NoStop}%
\bibitem [{\citenamefont {Motrunich}(2006)}]{motrunich2006orbital}%
  \BibitemOpen
  \bibfield  {author} {\bibinfo {author} {\bibfnamefont {O.~I.}\ \bibnamefont
  {Motrunich}},\ }\href@noop {} {\bibfield  {journal} {\bibinfo  {journal}
  {Physical Review B}\ }\textbf {\bibinfo {volume} {73}},\ \bibinfo {pages}
  {155115} (\bibinfo {year} {2006})}\BibitemShut {NoStop}%
\bibitem [{\citenamefont {Knolle}\ and\ \citenamefont
  {Moessner}(2019)}]{knolle2019field}%
  \BibitemOpen
  \bibfield  {author} {\bibinfo {author} {\bibfnamefont {J.}~\bibnamefont
  {Knolle}}\ and\ \bibinfo {author} {\bibfnamefont {R.}~\bibnamefont
  {Moessner}},\ }\href@noop {} {\bibfield  {journal} {\bibinfo  {journal}
  {Annual Review of Condensed Matter Physics}\ }\textbf {\bibinfo {volume}
  {10}},\ \bibinfo {pages} {451} (\bibinfo {year} {2019})}\BibitemShut
  {NoStop}%
\bibitem [{\citenamefont {Savary}\ and\ \citenamefont
  {Balents}(2016)}]{savary2016quantum}%
  \BibitemOpen
  \bibfield  {author} {\bibinfo {author} {\bibfnamefont {L.}~\bibnamefont
  {Savary}}\ and\ \bibinfo {author} {\bibfnamefont {L.}~\bibnamefont
  {Balents}},\ }\href@noop {} {\bibfield  {journal} {\bibinfo  {journal}
  {Reports on Progress in Physics}\ }\textbf {\bibinfo {volume} {80}},\
  \bibinfo {pages} {016502} (\bibinfo {year} {2016})}\BibitemShut {NoStop}%
\bibitem [{\citenamefont {Burch}\ \emph {et~al.}(2018)\citenamefont {Burch},
  \citenamefont {Mandrus},\ and\ \citenamefont {Park}}]{burch2018magnetism}%
  \BibitemOpen
  \bibfield  {author} {\bibinfo {author} {\bibfnamefont {K.~S.}\ \bibnamefont
  {Burch}}, \bibinfo {author} {\bibfnamefont {D.}~\bibnamefont {Mandrus}}, \
  and\ \bibinfo {author} {\bibfnamefont {J.-G.}\ \bibnamefont {Park}},\
  }\href@noop {} {\bibfield  {journal} {\bibinfo  {journal} {Nature}\ }\textbf
  {\bibinfo {volume} {563}},\ \bibinfo {pages} {47} (\bibinfo {year}
  {2018})}\BibitemShut {NoStop}%
\bibitem [{\citenamefont {Zhou}\ \emph {et~al.}(2019)\citenamefont {Zhou},
  \citenamefont {Balgley}, \citenamefont {Lampen-Kelley}, \citenamefont {Yan},
  \citenamefont {Mandrus},\ and\ \citenamefont {Henriksen}}]{zhou2019evidence}%
  \BibitemOpen
  \bibfield  {author} {\bibinfo {author} {\bibfnamefont {B.}~\bibnamefont
  {Zhou}}, \bibinfo {author} {\bibfnamefont {J.}~\bibnamefont {Balgley}},
  \bibinfo {author} {\bibfnamefont {P.}~\bibnamefont {Lampen-Kelley}}, \bibinfo
  {author} {\bibfnamefont {J.-Q.}\ \bibnamefont {Yan}}, \bibinfo {author}
  {\bibfnamefont {D.}~\bibnamefont {Mandrus}}, \ and\ \bibinfo {author}
  {\bibfnamefont {E.}~\bibnamefont {Henriksen}},\ }\href@noop {} {\bibfield
  {journal} {\bibinfo  {journal} {Physical Review B}\ }\textbf {\bibinfo
  {volume} {100}},\ \bibinfo {pages} {165426} (\bibinfo {year}
  {2019})}\BibitemShut {NoStop}%
\bibitem [{\citenamefont {Mashhadi}\ \emph {et~al.}(2019)\citenamefont
  {Mashhadi}, \citenamefont {Kim}, \citenamefont {Kim}, \citenamefont {Weber},
  \citenamefont {Taniguchi}, \citenamefont {Watanabe}, \citenamefont {Park},
  \citenamefont {Lotsch}, \citenamefont {Smet}, \citenamefont {Burghard},\ and\
  \citenamefont {Kern}}]{Mashhadi2019}%
  \BibitemOpen
  \bibfield  {author} {\bibinfo {author} {\bibfnamefont {S.}~\bibnamefont
  {Mashhadi}}, \bibinfo {author} {\bibfnamefont {Y.}~\bibnamefont {Kim}},
  \bibinfo {author} {\bibfnamefont {J.}~\bibnamefont {Kim}}, \bibinfo {author}
  {\bibfnamefont {D.}~\bibnamefont {Weber}}, \bibinfo {author} {\bibfnamefont
  {T.}~\bibnamefont {Taniguchi}}, \bibinfo {author} {\bibfnamefont
  {K.}~\bibnamefont {Watanabe}}, \bibinfo {author} {\bibfnamefont
  {N.}~\bibnamefont {Park}}, \bibinfo {author} {\bibfnamefont {B.}~\bibnamefont
  {Lotsch}}, \bibinfo {author} {\bibfnamefont {J.~H.}\ \bibnamefont {Smet}},
  \bibinfo {author} {\bibfnamefont {M.}~\bibnamefont {Burghard}}, \ and\
  \bibinfo {author} {\bibfnamefont {K.}~\bibnamefont {Kern}},\ }\href {\doibase
  10.1021/acs.nanolett.9b01691} {\bibfield  {journal} {\bibinfo  {journal}
  {Nano Letters}\ }\textbf {\bibinfo {volume} {19}},\ \bibinfo {pages}
  {4659–4665} (\bibinfo {year} {2019})}\BibitemShut {NoStop}%
\bibitem [{\citenamefont {Kitaev}(2006)}]{Kitaev2006}%
  \BibitemOpen
  \bibfield  {author} {\bibinfo {author} {\bibfnamefont {A.}~\bibnamefont
  {Kitaev}},\ }\href {\doibase 10.1016/j.aop.2005.10.005} {\bibfield  {journal}
  {\bibinfo  {journal} {Annals of Physics}\ }\textbf {\bibinfo {volume}
  {321}},\ \bibinfo {pages} {2–111} (\bibinfo {year} {2006})}\BibitemShut
  {NoStop}%
\bibitem [{\citenamefont {Rau}\ \emph {et~al.}(2016)\citenamefont {Rau},
  \citenamefont {Lee},\ and\ \citenamefont {Kee}}]{rau2016spin}%
  \BibitemOpen
  \bibfield  {author} {\bibinfo {author} {\bibfnamefont {J.~G.}\ \bibnamefont
  {Rau}}, \bibinfo {author} {\bibfnamefont {E.~K.-H.}\ \bibnamefont {Lee}}, \
  and\ \bibinfo {author} {\bibfnamefont {H.-Y.}\ \bibnamefont {Kee}},\
  }\href@noop {} {\  (\bibinfo {year} {2016})}\BibitemShut {NoStop}%
\bibitem [{\citenamefont {Hermanns}\ \emph {et~al.}(2018)\citenamefont
  {Hermanns}, \citenamefont {Kimchi},\ and\ \citenamefont
  {Knolle}}]{hermanns2018physics}%
  \BibitemOpen
  \bibfield  {author} {\bibinfo {author} {\bibfnamefont {M.}~\bibnamefont
  {Hermanns}}, \bibinfo {author} {\bibfnamefont {I.}~\bibnamefont {Kimchi}}, \
  and\ \bibinfo {author} {\bibfnamefont {J.}~\bibnamefont {Knolle}},\
  }\href@noop {} {\bibfield  {journal} {\bibinfo  {journal} {Annual Review of
  Condensed Matter Physics}\ }\textbf {\bibinfo {volume} {9}},\ \bibinfo
  {pages} {17} (\bibinfo {year} {2018})}\BibitemShut {NoStop}%
\bibitem [{\citenamefont {Sandilands}\ \emph {et~al.}(2015)\citenamefont
  {Sandilands}, \citenamefont {Tian}, \citenamefont {Plumb}, \citenamefont
  {Kim},\ and\ \citenamefont {Burch}}]{Sandilands2015}%
  \BibitemOpen
  \bibfield  {author} {\bibinfo {author} {\bibfnamefont {L.~J.}\ \bibnamefont
  {Sandilands}}, \bibinfo {author} {\bibfnamefont {Y.}~\bibnamefont {Tian}},
  \bibinfo {author} {\bibfnamefont {K.~W.}\ \bibnamefont {Plumb}}, \bibinfo
  {author} {\bibfnamefont {Y.-J.}\ \bibnamefont {Kim}}, \ and\ \bibinfo
  {author} {\bibfnamefont {K.~S.}\ \bibnamefont {Burch}},\ }\href {\doibase
  10.1103/PhysRevLett.114.147201} {\bibfield  {journal} {\bibinfo  {journal}
  {Phys. Rev. Lett.}\ }\textbf {\bibinfo {volume} {114}},\ \bibinfo {pages}
  {147201} (\bibinfo {year} {2015})}\BibitemShut {NoStop}%
\bibitem [{\citenamefont {Banerjee}\ \emph {et~al.}(2016)\citenamefont
  {Banerjee}, \citenamefont {Bridges}, \citenamefont {Yan}, \citenamefont
  {Aczel}, \citenamefont {Li}, \citenamefont {Stone}, \citenamefont {Granroth},
  \citenamefont {Lumsden}, \citenamefont {Yiu}, \citenamefont {Knolle},\ and\
  \citenamefont {et~al.}}]{Banerjee2016}%
  \BibitemOpen
  \bibfield  {author} {\bibinfo {author} {\bibfnamefont {A.}~\bibnamefont
  {Banerjee}}, \bibinfo {author} {\bibfnamefont {C.~A.}\ \bibnamefont
  {Bridges}}, \bibinfo {author} {\bibfnamefont {J.-Q.}\ \bibnamefont {Yan}},
  \bibinfo {author} {\bibfnamefont {A.~A.}\ \bibnamefont {Aczel}}, \bibinfo
  {author} {\bibfnamefont {L.}~\bibnamefont {Li}}, \bibinfo {author}
  {\bibfnamefont {M.~B.}\ \bibnamefont {Stone}}, \bibinfo {author}
  {\bibfnamefont {G.~E.}\ \bibnamefont {Granroth}}, \bibinfo {author}
  {\bibfnamefont {M.~D.}\ \bibnamefont {Lumsden}}, \bibinfo {author}
  {\bibfnamefont {Y.}~\bibnamefont {Yiu}}, \bibinfo {author} {\bibfnamefont
  {J.}~\bibnamefont {Knolle}}, \ and\ \bibinfo {author} {\bibnamefont
  {et~al.}},\ }\href {\doibase 10.1038/nmat4604} {\bibfield  {journal}
  {\bibinfo  {journal} {Nature Materials}\ }\textbf {\bibinfo {volume} {15}},\
  \bibinfo {pages} {733–740} (\bibinfo {year} {2016})}\BibitemShut {NoStop}%
\bibitem [{\citenamefont {Kasahara}\ \emph {et~al.}(2018)\citenamefont
  {Kasahara}, \citenamefont {Ohnishi}, \citenamefont {Mizukami}, \citenamefont
  {Tanaka}, \citenamefont {Ma}, \citenamefont {Sugii}, \citenamefont {Kurita},
  \citenamefont {Tanaka}, \citenamefont {Nasu}, \citenamefont {Motome},\ and\
  \citenamefont {et~al.}}]{Kasahara2018}%
  \BibitemOpen
  \bibfield  {author} {\bibinfo {author} {\bibfnamefont {Y.}~\bibnamefont
  {Kasahara}}, \bibinfo {author} {\bibfnamefont {T.}~\bibnamefont {Ohnishi}},
  \bibinfo {author} {\bibfnamefont {Y.}~\bibnamefont {Mizukami}}, \bibinfo
  {author} {\bibfnamefont {O.}~\bibnamefont {Tanaka}}, \bibinfo {author}
  {\bibfnamefont {S.}~\bibnamefont {Ma}}, \bibinfo {author} {\bibfnamefont
  {K.}~\bibnamefont {Sugii}}, \bibinfo {author} {\bibfnamefont
  {N.}~\bibnamefont {Kurita}}, \bibinfo {author} {\bibfnamefont
  {H.}~\bibnamefont {Tanaka}}, \bibinfo {author} {\bibfnamefont
  {J.}~\bibnamefont {Nasu}}, \bibinfo {author} {\bibfnamefont {Y.}~\bibnamefont
  {Motome}}, \ and\ \bibinfo {author} {\bibnamefont {et~al.}},\ }\href
  {\doibase 10.1038/s41586-018-0274-0} {\bibfield  {journal} {\bibinfo
  {journal} {Nature}\ }\textbf {\bibinfo {volume} {559}},\ \bibinfo {pages}
  {227–231} (\bibinfo {year} {2018})}\BibitemShut {NoStop}%
\bibitem [{\citenamefont {Winter}\ \emph {et~al.}(2017)\citenamefont {Winter},
  \citenamefont {Tsirlin}, \citenamefont {Daghofer}, \citenamefont {van~den
  Brink}, \citenamefont {Singh}, \citenamefont {Gegenwart},\ and\ \citenamefont
  {Valenti}}]{winter2017models}%
  \BibitemOpen
  \bibfield  {author} {\bibinfo {author} {\bibfnamefont {S.~M.}\ \bibnamefont
  {Winter}}, \bibinfo {author} {\bibfnamefont {A.~A.}\ \bibnamefont {Tsirlin}},
  \bibinfo {author} {\bibfnamefont {M.}~\bibnamefont {Daghofer}}, \bibinfo
  {author} {\bibfnamefont {J.}~\bibnamefont {van~den Brink}}, \bibinfo {author}
  {\bibfnamefont {Y.}~\bibnamefont {Singh}}, \bibinfo {author} {\bibfnamefont
  {P.}~\bibnamefont {Gegenwart}}, \ and\ \bibinfo {author} {\bibfnamefont
  {R.}~\bibnamefont {Valenti}},\ }\href@noop {} {\bibfield  {journal} {\bibinfo
   {journal} {Journal of Physics: Condensed Matter}\ }\textbf {\bibinfo
  {volume} {29}},\ \bibinfo {pages} {493002} (\bibinfo {year}
  {2017})}\BibitemShut {NoStop}%
\bibitem [{\citenamefont {Takagi}\ \emph {et~al.}(2019)\citenamefont {Takagi},
  \citenamefont {Takayama}, \citenamefont {Jackeli}, \citenamefont
  {Khaliullin},\ and\ \citenamefont {Nagler}}]{takagi2019concept}%
  \BibitemOpen
  \bibfield  {author} {\bibinfo {author} {\bibfnamefont {H.}~\bibnamefont
  {Takagi}}, \bibinfo {author} {\bibfnamefont {T.}~\bibnamefont {Takayama}},
  \bibinfo {author} {\bibfnamefont {G.}~\bibnamefont {Jackeli}}, \bibinfo
  {author} {\bibfnamefont {G.}~\bibnamefont {Khaliullin}}, \ and\ \bibinfo
  {author} {\bibfnamefont {S.~E.}\ \bibnamefont {Nagler}},\ }\href@noop {}
  {\bibfield  {journal} {\bibinfo  {journal} {Nature Reviews Physics}\ }\textbf
  {\bibinfo {volume} {1}},\ \bibinfo {pages} {264} (\bibinfo {year}
  {2019})}\BibitemShut {NoStop}%
\bibitem [{\citenamefont {Biswas}\ \emph {et~al.}(2019)\citenamefont {Biswas},
  \citenamefont {Li}, \citenamefont {Winter}, \citenamefont {Knolle},\ and\
  \citenamefont {Valent\'i}}]{DFT}%
  \BibitemOpen
  \bibfield  {author} {\bibinfo {author} {\bibfnamefont {S.}~\bibnamefont
  {Biswas}}, \bibinfo {author} {\bibfnamefont {Y.}~\bibnamefont {Li}}, \bibinfo
  {author} {\bibfnamefont {S.~M.}\ \bibnamefont {Winter}}, \bibinfo {author}
  {\bibfnamefont {J.}~\bibnamefont {Knolle}}, \ and\ \bibinfo {author}
  {\bibfnamefont {R.}~\bibnamefont {Valent\'i}},\ }\href {\doibase
  10.1103/PhysRevLett.123.237201} {\bibfield  {journal} {\bibinfo  {journal}
  {Phys. Rev. Lett.}\ }\textbf {\bibinfo {volume} {123}},\ \bibinfo {pages}
  {237201} (\bibinfo {year} {2019})}\BibitemShut {NoStop}%
\bibitem [{\citenamefont {Wang}\ \emph {et~al.}(2020)\citenamefont {Wang},
  \citenamefont {Balgley}, \citenamefont {Gerber}, \citenamefont {Gray},
  \citenamefont {Kumar}, \citenamefont {Lu}, \citenamefont {Yan}, \citenamefont
  {Fereidouni}, \citenamefont {Basnet}, \citenamefont {Yun} \emph
  {et~al.}}]{wang2020modulation}%
  \BibitemOpen
  \bibfield  {author} {\bibinfo {author} {\bibfnamefont {Y.}~\bibnamefont
  {Wang}}, \bibinfo {author} {\bibfnamefont {J.}~\bibnamefont {Balgley}},
  \bibinfo {author} {\bibfnamefont {E.}~\bibnamefont {Gerber}}, \bibinfo
  {author} {\bibfnamefont {M.}~\bibnamefont {Gray}}, \bibinfo {author}
  {\bibfnamefont {N.}~\bibnamefont {Kumar}}, \bibinfo {author} {\bibfnamefont
  {X.}~\bibnamefont {Lu}}, \bibinfo {author} {\bibfnamefont {J.-Q.}\
  \bibnamefont {Yan}}, \bibinfo {author} {\bibfnamefont {A.}~\bibnamefont
  {Fereidouni}}, \bibinfo {author} {\bibfnamefont {R.}~\bibnamefont {Basnet}},
  \bibinfo {author} {\bibfnamefont {S.~J.}\ \bibnamefont {Yun}},  \emph
  {et~al.},\ }\href@noop {} {\bibfield  {journal} {\bibinfo  {journal} {arXiv
  preprint arXiv:2007.06603}\ } (\bibinfo {year} {2020})}\BibitemShut {NoStop}%
\bibitem [{\citenamefont {Gerber}\ \emph {et~al.}(2020)\citenamefont {Gerber},
  \citenamefont {Yao}, \citenamefont {Arias},\ and\ \citenamefont
  {Kim}}]{gerber2020ab}%
  \BibitemOpen
  \bibfield  {author} {\bibinfo {author} {\bibfnamefont {E.}~\bibnamefont
  {Gerber}}, \bibinfo {author} {\bibfnamefont {Y.}~\bibnamefont {Yao}},
  \bibinfo {author} {\bibfnamefont {T.~A.}\ \bibnamefont {Arias}}, \ and\
  \bibinfo {author} {\bibfnamefont {E.-A.}\ \bibnamefont {Kim}},\ }\href@noop
  {} {\bibfield  {journal} {\bibinfo  {journal} {Physical Review Letters}\
  }\textbf {\bibinfo {volume} {124}},\ \bibinfo {pages} {106804} (\bibinfo
  {year} {2020})}\BibitemShut {NoStop}%
\bibitem [{\citenamefont {Seifert}\ \emph {et~al.}(2018)\citenamefont
  {Seifert}, \citenamefont {Meng},\ and\ \citenamefont {Vojta}}]{Meng2018}%
  \BibitemOpen
  \bibfield  {author} {\bibinfo {author} {\bibfnamefont {U.~F.~P.}\
  \bibnamefont {Seifert}}, \bibinfo {author} {\bibfnamefont {T.}~\bibnamefont
  {Meng}}, \ and\ \bibinfo {author} {\bibfnamefont {M.}~\bibnamefont {Vojta}},\
  }\href {\doibase 10.1103/physrevb.97.085118} {\bibfield  {journal} {\bibinfo
  {journal} {Physical Review B}\ }\textbf {\bibinfo {volume} {97}} (\bibinfo
  {year} {2018}),\ 10.1103/physrevb.97.085118}\BibitemShut {NoStop}%
\bibitem [{\citenamefont {Choi}\ \emph {et~al.}(2018)\citenamefont {Choi},
  \citenamefont {Klein}, \citenamefont {Rosch},\ and\ \citenamefont
  {Kim}}]{choi2018topological}%
  \BibitemOpen
  \bibfield  {author} {\bibinfo {author} {\bibfnamefont {W.}~\bibnamefont
  {Choi}}, \bibinfo {author} {\bibfnamefont {P.~W.}\ \bibnamefont {Klein}},
  \bibinfo {author} {\bibfnamefont {A.}~\bibnamefont {Rosch}}, \ and\ \bibinfo
  {author} {\bibfnamefont {Y.~B.}\ \bibnamefont {Kim}},\ }\href@noop {}
  {\bibfield  {journal} {\bibinfo  {journal} {Physical Review B}\ }\textbf
  {\bibinfo {volume} {98}},\ \bibinfo {pages} {155123} (\bibinfo {year}
  {2018})}\BibitemShut {NoStop}%
\bibitem [{\citenamefont {Burnell}\ and\ \citenamefont
  {Nayak}(2011)}]{Burnell2011}%
  \BibitemOpen
  \bibfield  {author} {\bibinfo {author} {\bibfnamefont {F.~J.}\ \bibnamefont
  {Burnell}}\ and\ \bibinfo {author} {\bibfnamefont {C.}~\bibnamefont
  {Nayak}},\ }\href {\doibase 10.1103/PhysRevB.84.125125} {\bibfield  {journal}
  {\bibinfo  {journal} {Phys. Rev. B}\ }\textbf {\bibinfo {volume} {84}},\
  \bibinfo {pages} {125125} (\bibinfo {year} {2011})}\BibitemShut {NoStop}%
\bibitem [{\citenamefont {Schaffer}\ \emph {et~al.}(2012)\citenamefont
  {Schaffer}, \citenamefont {Bhattacharjee},\ and\ \citenamefont
  {Kim}}]{Schaffer2012}%
  \BibitemOpen
  \bibfield  {author} {\bibinfo {author} {\bibfnamefont {R.}~\bibnamefont
  {Schaffer}}, \bibinfo {author} {\bibfnamefont {S.}~\bibnamefont
  {Bhattacharjee}}, \ and\ \bibinfo {author} {\bibfnamefont {Y.~B.}\
  \bibnamefont {Kim}},\ }\href {\doibase 10.1103/PhysRevB.86.224417} {\bibfield
   {journal} {\bibinfo  {journal} {Phys. Rev. B}\ }\textbf {\bibinfo {volume}
  {86}},\ \bibinfo {pages} {224417} (\bibinfo {year} {2012})}\BibitemShut
  {NoStop}%
\bibitem [{\citenamefont {Knolle}\ \emph {et~al.}(2018)\citenamefont {Knolle},
  \citenamefont {Bhattacharjee},\ and\ \citenamefont
  {Moessner}}]{knolle2018dynamics}%
  \BibitemOpen
  \bibfield  {author} {\bibinfo {author} {\bibfnamefont {J.}~\bibnamefont
  {Knolle}}, \bibinfo {author} {\bibfnamefont {S.}~\bibnamefont
  {Bhattacharjee}}, \ and\ \bibinfo {author} {\bibfnamefont {R.}~\bibnamefont
  {Moessner}},\ }\href@noop {} {\bibfield  {journal} {\bibinfo  {journal}
  {Physical Review B}\ }\textbf {\bibinfo {volume} {97}},\ \bibinfo {pages}
  {134432} (\bibinfo {year} {2018})}\BibitemShut {NoStop}%
\bibitem [{\citenamefont {Castro~Neto}\ \emph {et~al.}(2009)\citenamefont
  {Castro~Neto}, \citenamefont {Guinea}, \citenamefont {Peres}, \citenamefont
  {Novoselov},\ and\ \citenamefont {Geim}}]{Castro2009}%
  \BibitemOpen
  \bibfield  {author} {\bibinfo {author} {\bibfnamefont {A.~H.}\ \bibnamefont
  {Castro~Neto}}, \bibinfo {author} {\bibfnamefont {F.}~\bibnamefont {Guinea}},
  \bibinfo {author} {\bibfnamefont {N.~M.~R.}\ \bibnamefont {Peres}}, \bibinfo
  {author} {\bibfnamefont {K.~S.}\ \bibnamefont {Novoselov}}, \ and\ \bibinfo
  {author} {\bibfnamefont {A.~K.}\ \bibnamefont {Geim}},\ }\href {\doibase
  10.1103/RevModPhys.81.109} {\bibfield  {journal} {\bibinfo  {journal} {Rev.
  Mod. Phys.}\ }\textbf {\bibinfo {volume} {81}},\ \bibinfo {pages} {109}
  (\bibinfo {year} {2009})}\BibitemShut {NoStop}%
\bibitem [{\citenamefont {Hartnoll}\ and\ \citenamefont
  {Hofman}(2010)}]{GLK_Hartnoll_Hofman}%
  \BibitemOpen
  \bibfield  {author} {\bibinfo {author} {\bibfnamefont {S.~A.}\ \bibnamefont
  {Hartnoll}}\ and\ \bibinfo {author} {\bibfnamefont {D.~M.}\ \bibnamefont
  {Hofman}},\ }\href {\doibase 10.1103/physrevb.81.155125} {\bibfield
  {journal} {\bibinfo  {journal} {Physical Review B}\ }\textbf {\bibinfo
  {volume} {81}} (\bibinfo {year} {2010}),\
  10.1103/physrevb.81.155125}\BibitemShut {NoStop}%
\bibitem [{\citenamefont {Banerjee}\ \emph {et~al.}(2018)\citenamefont
  {Banerjee}, \citenamefont {Lampen-Kelley}, \citenamefont {Knolle},
  \citenamefont {Balz}, \citenamefont {Aczel}, \citenamefont {Winn},
  \citenamefont {Liu}, \citenamefont {Pajerowski}, \citenamefont {Yan},
  \citenamefont {Bridges} \emph {et~al.}}]{banerjee2018excitations}%
  \BibitemOpen
  \bibfield  {author} {\bibinfo {author} {\bibfnamefont {A.}~\bibnamefont
  {Banerjee}}, \bibinfo {author} {\bibfnamefont {P.}~\bibnamefont
  {Lampen-Kelley}}, \bibinfo {author} {\bibfnamefont {J.}~\bibnamefont
  {Knolle}}, \bibinfo {author} {\bibfnamefont {C.}~\bibnamefont {Balz}},
  \bibinfo {author} {\bibfnamefont {A.~A.}\ \bibnamefont {Aczel}}, \bibinfo
  {author} {\bibfnamefont {B.}~\bibnamefont {Winn}}, \bibinfo {author}
  {\bibfnamefont {Y.}~\bibnamefont {Liu}}, \bibinfo {author} {\bibfnamefont
  {D.}~\bibnamefont {Pajerowski}}, \bibinfo {author} {\bibfnamefont
  {J.}~\bibnamefont {Yan}}, \bibinfo {author} {\bibfnamefont {C.~A.}\
  \bibnamefont {Bridges}},  \emph {et~al.},\ }\href@noop {} {\bibfield
  {journal} {\bibinfo  {journal} {npj Quantum Materials}\ }\textbf {\bibinfo
  {volume} {3}},\ \bibinfo {pages} {1} (\bibinfo {year} {2018})}\BibitemShut
  {NoStop}%
\bibitem [{\citenamefont {Rizzo}\ \emph {et~al.}(2020)\citenamefont {Rizzo},
  \citenamefont {Jessen}, \citenamefont {Sun}, \citenamefont {Ruta},
  \citenamefont {Zhang}, \citenamefont {Yan}, \citenamefont {Xian},
  \citenamefont {McLeod}, \citenamefont {Berkowitz}, \citenamefont {Watanabe}
  \emph {et~al.}}]{rizzo2020graphene}%
  \BibitemOpen
  \bibfield  {author} {\bibinfo {author} {\bibfnamefont {D.~J.}\ \bibnamefont
  {Rizzo}}, \bibinfo {author} {\bibfnamefont {B.~S.}\ \bibnamefont {Jessen}},
  \bibinfo {author} {\bibfnamefont {Z.}~\bibnamefont {Sun}}, \bibinfo {author}
  {\bibfnamefont {F.~L.}\ \bibnamefont {Ruta}}, \bibinfo {author}
  {\bibfnamefont {J.}~\bibnamefont {Zhang}}, \bibinfo {author} {\bibfnamefont
  {J.-Q.}\ \bibnamefont {Yan}}, \bibinfo {author} {\bibfnamefont
  {L.}~\bibnamefont {Xian}}, \bibinfo {author} {\bibfnamefont {A.~S.}\
  \bibnamefont {McLeod}}, \bibinfo {author} {\bibfnamefont {M.~E.}\
  \bibnamefont {Berkowitz}}, \bibinfo {author} {\bibfnamefont {K.}~\bibnamefont
  {Watanabe}},  \emph {et~al.},\ }\href@noop {} {\bibfield  {journal} {\bibinfo
   {journal} {arXiv preprint arXiv:2007.07147}\ } (\bibinfo {year}
  {2020})}\BibitemShut {NoStop}%
\bibitem [{\citenamefont {Law}\ and\ \citenamefont {Lee}(2017)}]{law20171t}%
  \BibitemOpen
  \bibfield  {author} {\bibinfo {author} {\bibfnamefont {K.~T.}\ \bibnamefont
  {Law}}\ and\ \bibinfo {author} {\bibfnamefont {P.~A.}\ \bibnamefont {Lee}},\
  }\href@noop {} {\bibfield  {journal} {\bibinfo  {journal} {Proceedings of the
  National Academy of Sciences}\ }\textbf {\bibinfo {volume} {114}},\ \bibinfo
  {pages} {6996} (\bibinfo {year} {2017})}\BibitemShut {NoStop}%
\bibitem [{\citenamefont {Pasco}\ \emph {et~al.}(2019)\citenamefont {Pasco},
  \citenamefont {El~Baggari}, \citenamefont {Bianco}, \citenamefont
  {Kourkoutis},\ and\ \citenamefont {McQueen}}]{pasco2019tunable}%
  \BibitemOpen
  \bibfield  {author} {\bibinfo {author} {\bibfnamefont {C.~M.}\ \bibnamefont
  {Pasco}}, \bibinfo {author} {\bibfnamefont {I.}~\bibnamefont {El~Baggari}},
  \bibinfo {author} {\bibfnamefont {E.}~\bibnamefont {Bianco}}, \bibinfo
  {author} {\bibfnamefont {L.~F.}\ \bibnamefont {Kourkoutis}}, \ and\ \bibinfo
  {author} {\bibfnamefont {T.~M.}\ \bibnamefont {McQueen}},\ }\href@noop {}
  {\bibfield  {journal} {\bibinfo  {journal} {ACS nano}\ }\textbf {\bibinfo
  {volume} {13}},\ \bibinfo {pages} {9457} (\bibinfo {year}
  {2019})}\BibitemShut {NoStop}%
\end{thebibliography}
%

\clearpage
\appendix

\begin{widetext}
    \begin{center}
      {{\bf Supplementary Material:}}
    \end{center}
    \end{widetext}

\section{Landau level structure}
We introduce a magnetic field $B$ over the vector potential $\v{A} = (-B_z y, -B_x z, -B_y x)^\mathsf{T}$ given in the Landau-gauge. Within minimal coupling this can be described by extending the kinetic momentum to the canonical momentum $\hbar \v{q} \rightarrow \hbar \v{q}- e\v{A}$. Since we are considering a two dimensional model only the out-of-plane component of the $B$-field couples to momentum. Note that the Landau gauge breaks the translational symmetry in $y$-direction such that $q_y$ is not a good quantum number anymore.

Analogously to determining the Landau levels in graphene, momentum dependent entries correspond to ladder operators of the harmonic oscillator, such that the LL-Hamiltonian reads

\begin{align}
H^K =\sum_{q_x, \sigma}  
\v{\Psi}_{0,q_x,\sigma}^{K \dagger}
&\begin{pmatrix}
0 & 0 & 0 &0 \\
0 &W&0&\frac{J}{2} \\
0 &0& 0& 0\\
0& \frac{J}{2} & 0&0\\
\end{pmatrix}
\v{\Psi}_{0,q_x,\sigma}^{K} \nonumber\\
 + \sum_{l=1,q_x, \sigma}  
\v{\Psi}_{l,q_x,\sigma}^{K \dagger}
&\begin{pmatrix}
W & \omega_t \sqrt{l} & \frac{J}{2} &0 \\
\omega_t \sqrt{l} &W&0&\frac{J}{2} \\
\frac{J}{2} &0& 0& \omega_K \sqrt{l}\\
0& \frac{J}{2} & \omega_K \sqrt{l}&0\\
\end{pmatrix}
\v{\Psi}_{l,q_x,\sigma}^{K}.
\label{eq:H_K_LL}
\end{align}
Diagonalizing \eqref{eq:H_K_LL} results in Eq.\eqref{LL} of the main text where the zeroth LL $l=0$ has an inherently different form but is not of interested for our current work.

\section{Extending the generalized Lifshitz-Kosevich formula}
\label{sec:extending_GLK}
The goal is to extend the calculation done from (27) to (25) in \cite{GLK_Hartnoll_Hofman} to poles where $\im l^\star <0$ or $\re l^\star <0$. (27) in \cite{GLK_Hartnoll_Hofman} reads
\begin{equation}
\Omega_\text{osc.} = -N_\phi T \re \sum_{n=0}^\infty \sum_{l=0}^\infty \ln(l-l^\star_n).
\end{equation}
By showing that, up to non-oscillatory terms, the fundamental formula
\begin{equation}
\sum_{l=0}^\infty \ln(l-l^\star) = -\theta(\re l^\star)\sum_{k=1}^\infty \frac{1}{k} \e^{2 \pi \ii k \l^\star \mathrm{sgn}(\im l^\star)} 
\label{eq:GLK_fundamental_connection}
\end{equation}
holds, $\Omega_\text{osc.}$ can be simplified to \eqref{GLK}.

To proof \eqref{eq:GLK_fundamental_connection} we follow the brief description given in \cite{GLK_Hartnoll_Hofman}. First we perform a Poisson resummation as given in (14) in \cite{GLK_Hartnoll_Hofman} and introduce $k \rightarrow k+\ii \epsilon$ with $\epsilon \rightarrow 0^+$ to ensure convergence. Then we integrate by parts and split up the sum over $k$.
\begin{widetext}
\begin{align}
& \sum_{l=0}^\infty \ln(l-l^\star) = \sum_{k=-\infty}^\infty \int_{0^-}^\infty \ln(x-l^\star) \e^{2 \pi \ii (k+\ii \epsilon) x} \d x\nonumber \\
&= \frac{\ln -l^\star}{2 \pi \epsilon}+\frac{1}{2 \pi \epsilon} \int_{0^-}^\infty \frac{1}{x-l^\star} \e^{-2 \pi \epsilon x} \d x-\frac{\ln -l^\star}{2 \pi \ii} \sum_{k=1}^\infty \left(\frac{1}{k+\ii \epsilon}+\frac{1}{-k+\ii \epsilon} \right) -\sum_{k=1}^\infty \frac{1}{2 \pi \ii k} \int_{0^-}^\infty \left(\e^{2 \pi \ii k x} -\e^{-2 \pi \ii k x} \right) \frac{\e^{-2 \pi \epsilon x}}{x-l^\star} \d x \nonumber \\
&= -\sum_{k=1}^\infty \frac{1}{2 \pi \ii k} \int_{0^-}^\infty \left(\e^{2 \pi \ii k x} -\e^{-2 \pi \ii k x} \right) \frac{\e^{-2 \pi \epsilon x}}{x-l^\star} \d x
\label{eq:GLK_appendix1}
\end{align}
\end{widetext}
We have dropped terms which will become large if we take $\epsilon \rightarrow 0^+$ but are non-oscillating. The remaining term consists of two exponentials with different signs in the exponent which can be evaluated separately by using complex contour integration. For the first exponential we choose a path $\gamma^+_1$ stretching from $0$ to $\infty$, then a path $\gamma^+_{\text{arc}}$ from $\infty$ to $\ii \infty$ and then along  $\gamma^+_2$ back to 0. For the second exponential these paths are mirrored at the abscissa.

In both cases we can rewrite, using the residue theorem, the integral along $\gamma^\pm_1$ in terms of the integral along $\gamma^\pm_{\text{arc}}$ and $\gamma^\pm_{2}$ and a contribution of the poles in the first ($+$) or fourth ($-$) quadrant. Since the integrand vanishes for $x \rightarrow \infty$ the integral along $\gamma^\pm_{\text{arc}}$ vanishes. Furthermore the integral along $\gamma^\pm_{2}$ is non-oscillatory.
%Calculation showing that this is non oscillatory
%\begin{align}
%\int_{\gamma^+_2} \frac{\e^{2 \pi \ii (k+\ii\epsilon) x}}{x-l^\star} \d x 
%= \int_{\ii \infty}^0 \frac{\e^{2 \pi \ii k x}}{x-l^\star} \d x  
%= -\ii \int_{0}^{\infty} \frac{\e^{-2 \pi k \tilde{x}}}{\ii \tilde{x}-l^\star} \d \tilde{x}. 
%\end{align}
Therefore the only oscillatory contribution of the integral comes from the residue of the poles, located in the right half plane
\begin{align}
& \frac{1}{2 \pi \ii k} \int_{0^-}^\infty \left(\e^{2 \pi \ii k x} -\e^{-2 \pi \ii k x} \right) \frac{1}{x-l^\star} \d x \nonumber \\
&=\frac{1}{k} \theta(\re l^\star) \e^{2 \pi \ii k l^\star \mathrm{sgn}(\im l^\star)}.
\label{eq:GLK_appendix2}
\end{align}

\section{Total amplitude behavior}
Here, we would like to explore how the total amplitude of anomalous QOs depends on microscopic parameters. The maximum of $R(T)$ from \eqref{R} is roughly located at $T_\text{max} \approx \nicefrac{J}{5}$. The decay of $R(\nicefrac{J}{5})$ is dominated by the zeroth Matsubara frequency, all other terms lead to a stronger decay. As a result we find that $R(\nicefrac{J}{5})$ is damped exponentially with the size of the hybridization region $J$

\begin{equation}
\exp \left( - \kappa \frac{J W}{\omega_t^2}\right)
\end{equation}
where $\kappa$ is a constant of order 1. This predicts that the QOs should only be observable up to a hybridization energy scale $J_\text{max} \sim \nicefrac{\omega_t^2}{W}$. For the graphene/ \RuCl heterostructure $J_\text{max} \sim \unit[1.6]{meV} \times \unit[B]{[T]}$ which is already for small magnetic fields sufficiently higher than our predicted values of $J\approx\unit[2]{meV}$.
Note, this estimate holds for the the linear Dirac band structure in the graphene/ \RuCl system. In systems with quadratic bands \cite{Knolle_Cooper_QOwithoutFS} $J_\text{max} \sim \omega_c$ such that much smaller hybridization scales are needed to observe the anomalous QOs.

\section{Additional fits, fitting parameters and Fourier spectra}

\begin{figure}[h]
	\centering
	\includegraphics[width=\columnwidth]{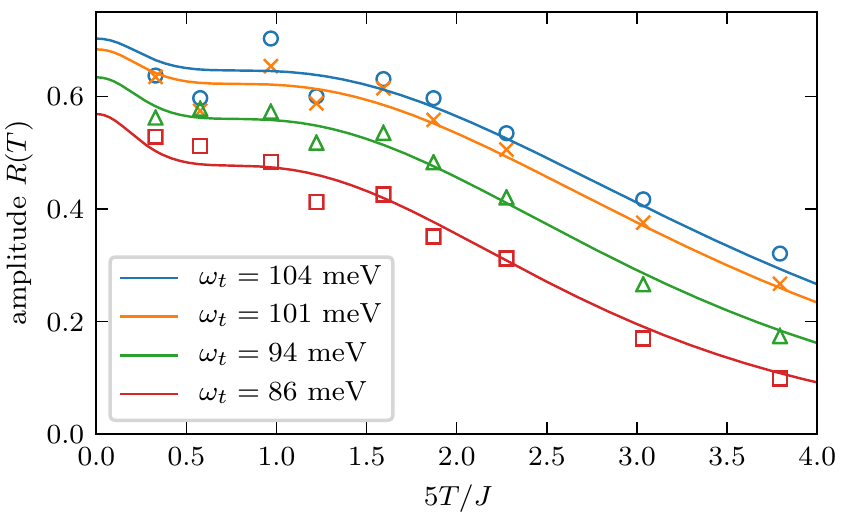}
		\caption{Fit of various amplitudes at different $B$-fields for sample B, predicting $|\mu| = \unit[0.41]{meV}$ and $J = \unit[1.71]{meV}$}
	\label{fig:R_diffB}
\end{figure}

\begin{figure}[h]
	\centering
	\includegraphics[width=\columnwidth]{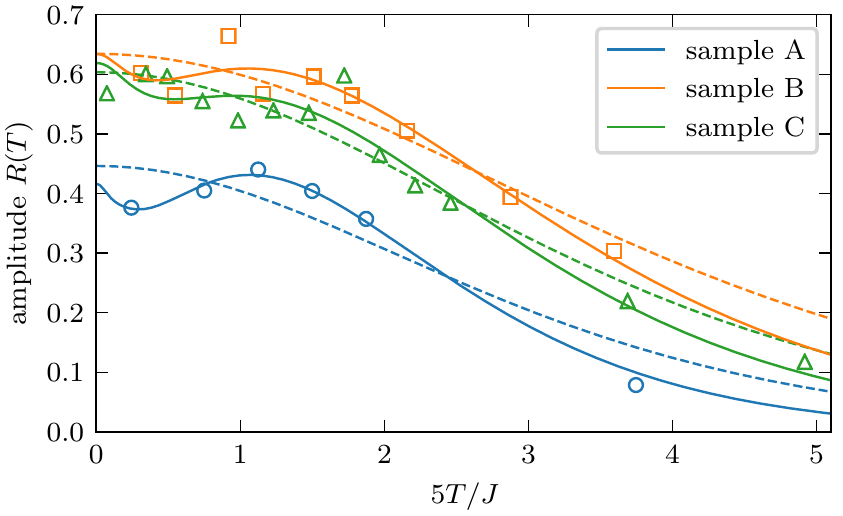}
		\caption{Same as Fig.\ref{fig3}~(b) but $\omega_t$ is now a fitting parameter for the LK-fits. The LK-fits have improved with respect to Fig.\ref{fig3}~(b) but the values for $\omega_t$, see Tab.II, are at odds with the microscopic values of $\omega_t$ which are derived from the DFT band structure and the corresponding magnetic field.}
	\label{fig:R_withwt}
\end{figure}

\begin{table}[!h]
	\centering
	\begin{tabular}{c|c c c c c}
		sample &A&B&C& B  for various $B$s\\
		\hline
		$f_{B^{-1}}$&$\unit[365]{T}$&$\unit[390]{T}$&$\unit[386]{T}$ &$\unit[390]{T}$\\
		$B_z$&$\unit[10]{T}$&$\unit[11.7]{T}$&$\unit[10]{T}$&$\unit[8/9.5/11/11.7]{T}$\\
		$W$&$\unit[583]{meV}$&$\unit[603]{meV}$&$\unit[600]{meV}$&$\unit[603]{meV}$\\
		$\omega_t$&$\unit[96.5]{meV}$&$\unit[104]{meV}$&$\unit[96.5]{meV}$&$\unit[86/94/101/104]{meV}$\\
		$J$&$\unit[2.30]{meV}$&$\unit[1.80]{meV}$&$\unit[1.75]{meV}$&$\unit[1.70]{meV}$\\
		$|\mu |$&$\unit[271]{\mu eV}$&$\unit[308]{\mu eV}$&$\unit[363]{\mu eV}$&$\unit[407]{\mu eV}$\\
		$R^2$ & 0.99917&0.99770&0.99505&-\\
		Adj.$R^2$ &  0.99834&0.99655&0.99357&-\\
	\end{tabular}
	\caption{Fitting parameters and quality criteria of the solid curves in Fig.\ref{fig3}~(b) (first three columns) and Fig.\ref{fig:R_diffB} (last column). $W$ and $\omega_t$ are calculated from the magnetic field $B$ and the frequency $f_{B^{-1}}$ which are determined from the experiment, $J$ and $\mu$ are fitting parameters. As quality criteria we show R-squared and Adjusted R-squared in the last two rows. }
	\label{tab:Param_DampingFac}
\end{table}

\begin{table}[!h]
	\centering
	\begin{tabular}{c|c c c c}
		sample & $W$ & $\omega_t$ & $R^2$ & Adj.$R^2$\\
		\hline
		A &$\unit[583]{meV}$&$\unit[96.5]{meV}$&0.95517&0.94620\\
		B &$\unit[603]{meV}$&$\unit[104]{meV}$&0.98497&0.98309\\
		C &$\unit[600]{meV}$&$\unit[96.5]{meV}$&0.98317&0.98177\\
		\hline
		A &$\unit[583]{meV}$&$\unit[117]{meV}$&0.98201&0.97301\\
		B &$\unit[603]{meV}$&$\unit[121]{meV}$&0.99548&0.99418\\
		C &$\unit[600]{meV}$&$\unit[110]{meV}$&0.99420&0.99315\\
	\end{tabular}
	\caption{(Fitting) parameters and quality criteria of the LK curves (dashed) in Fig.\ref{fig3}~(b) (first three rows) and Fig.\ref{fig:R_withwt} (last three rows).}
	\label{tab:Param_LKfits}
\end{table}
The value of Adj. $R^2$ (Adjusted $R^2$) accounts for the larger number of fitting parameters of the non-LK fit, compared to the LK-fit. For calculation we have used the in-build Mathematica routine.

\clearpage
\begin{figure*}[!h]
	\centering
	\includegraphics[width=\textwidth]{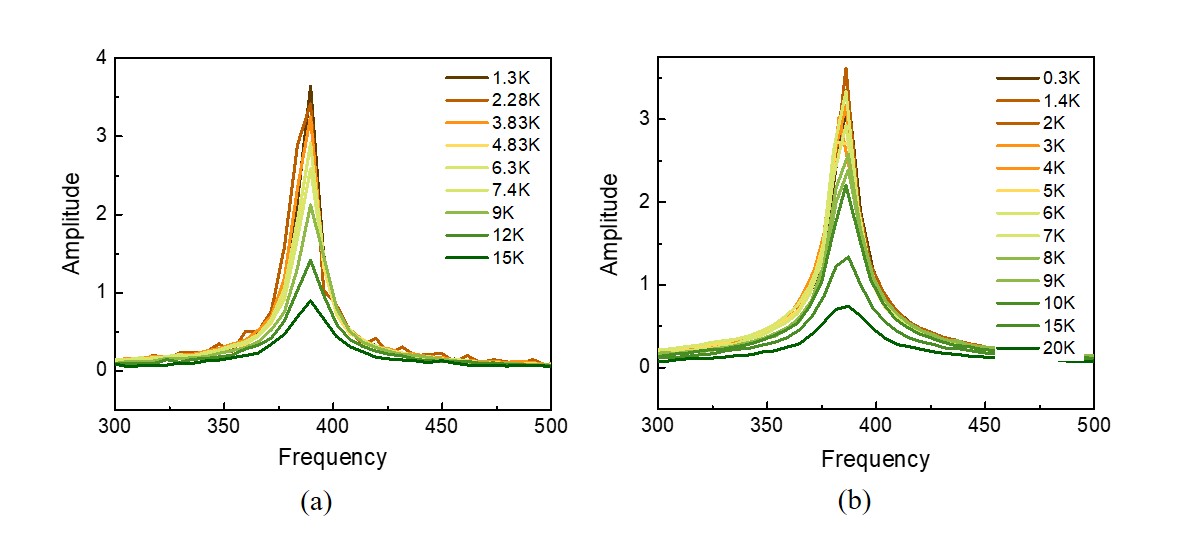}
	\caption{Fourier spectra in $B^{-1}$ of the two samples B (a) and C (b) for various temperatures, predicting clear maxima at $f_{B^{-1}}(B) = \unit[390]{T}$ and $f_{B^{-1}}(C) = \unit[386]{T}$.}
	\label{fig:FFTs}
\end{figure*}

%\section{Additional experimental data and effect of in-plane magnetic field}

\end{document}